\documentclass[12pt]{article}
\usepackage[dvips]{graphics}
\usepackage{pstricks,pst-plot}
\usepackage{amsmath,amssymb,mathrsfs}

\newlength{\dinwidth}
\newlength{\dinmargin}
\begin{document}

\renewcommand{\square}{\vrule height 1.5ex width 1.2ex depth -.1ex }


\newcommand{\II}{\leavevmode\hbox{\rm{\small1\kern-3.8pt\normalsize1}}}

\newcommand{\CC}{{\mathbb C}}
\newcommand{\RR}{{\mathbb R}}
\newcommand{\NN}{{\mathbb N}}
\newcommand{\QQ}{{\mathbb Q}}
\newcommand{\ZZ}{{\mathbb Z}}

\newcommand{\erf}{\mathop{\rm erf}}
\newcommand{\eps}{\varepsilon}
\newcommand{\esup}{\mathop{\rm ess\,sup}}


\newcommand{\CoinfM}{C_0^\infty(M)}
\newcommand{\CoinfN}{C_0^\infty(N)}
\newcommand{\Coinfd}{C_0^\infty(\RR^d\backslash\{ 0\})}
\newcommand{\Coinf}[1]{C_0^\infty(\RR^{#1}\backslash\{ 0\})}
\newcommand{\CoinX}[1]{C_0^\infty({#1})}
\newcommand{\Coin}{C_0^\infty(0,\infty)}


\newtheorem{Thm}{Theorem}[section]
\newtheorem{Def}[Thm]{Definition}
\newtheorem{Lem}[Thm]{Lemma}
\newtheorem{Prop}[Thm]{Proposition}
\newtheorem{Cor}[Thm]{Corollary}

\numberwithin{equation}{section}



\newcommand{\DD}{{\mathscr D}}
\newcommand{\EE}{{\mathscr E}}
\newcommand{\HH}{{\mathscr H}}
\newcommand{\KK}{{\mathscr K}}
\newcommand{\FF}{{\mathscr F}}
\newcommand{\Sch}{{\mathscr S}}

\newcommand{\DDco}{{\mathscr D}_{\rm cosp}}
\newcommand{\Aa}{{\cal A}}
\newcommand{\Ff}{{\cal F}}
\newcommand{\Uu}{{\cal U}}
\newcommand{\Xx}{{\cal X}}
\newcommand{\Zz}{{\cal Z}}
\newcommand{\Ft}{{\widetilde{\Ff}}}

\newcommand{\gb}{{\boldsymbol{g}}}
\newcommand{\hb}{{\boldsymbol{h}}}
\newcommand{\jb}{{\boldsymbol{j}}}
\newcommand{\kb}{{\boldsymbol{k}}}
\newcommand{\nb}{{\boldsymbol{n}}}
\newcommand{\xb}{{\boldsymbol{x}}}
\newcommand{\Fb}{{\boldsymbol{F}}}
\newcommand{\xbo}{{\boldsymbol{x_0}}}
\newcommand{\etb}{{\boldsymbol{\eta}}}


\newcommand{\Wf}{{\mathfrak W}}
\newcommand{\Af}{{\mathfrak{A}}}

\newcommand{\Dal}{\fbox{\phantom{${\scriptstyle *}$}}}

\newcommand{\Ran}{{\rm Ran}\,}
\newcommand{\supp}{{\rm supp}\,}
\newcommand{\Span}{{\rm span}\,}
\newcommand{\Tr}{{\rm Tr}\,}
\renewcommand{\Re}{{\rm Re}\,}
\renewcommand{\Im}{{\rm Im}\,}

\newcommand{\ip}[2]{{\langle #1\mid #2\rangle}}
\newcommand{\ket}[1]{{\mid #1\rangle}}
\newcommand{\bra}[1]{{\langle #1 \mid}}
\newcommand{\stack}[2]{\substack{#1 \\ #2}}

\newcommand{\ub}{\overline{u}}
\newcommand{\vb}{\overline{v}}
\newcommand{\wb}{\overline{w}}
\newcommand{\Bb}{\overline{B}}
\newcommand{\Pb}{\overline{P}}
\newcommand{\Qb}{\overline{Q}}
\newcommand{\Xb}{\overline{X}}
\newcommand{\Xib}{\overline{\Xi}}
\newcommand{\Xxb}{\overline{\Xx}}
\newcommand{\Yb}{\overline{Y}}

\newcommand{\kt}{\widetilde{k}}

\newcommand{\fhat}{\widehat{f}}
\newcommand{\hhat}{\widehat{h}}

\newcommand{\WF}{{\rm WF}\,}

\newcommand{\Had}{{\rm\sf Had}\,}

\newcommand{\LL}{{\mathcal L}}

\newcommand{\omABV}{\omega_{2AB(V)}}
\newcommand{\omABVl}{\omega^{\rm loc}_{2AB(V)}}

\newcommand{\Rr}{{\mathscr R}}

\newcommand{\spec}{{\rm spec}\,}

\newcommand{\OO}{{\mathscr O}}

\newcommand{\hs}{{\rm H.S.}}

\noindent
\begin{center}
{ \Large \bf Quantum Inequalities in Quantum Mechanics}
\\[35pt]
{\large \sc Simon P.\ Eveson${}^{1\dagger}$,\ \
Christopher J.\ Fewster${}^{1\ast}$\ \\[6pt] 
 {\rm and}\ \   Rainer Verch${}^2$}
\\[20pt]  
                 ${}^1$ Department of Mathematics,\\
                 University of York,\\
                 Heslington,\\
                 York YO10 5DD, United Kingdom\\[4pt]
                 ${}^{\dagger}$ e-mail: spe1$@$york.ac.uk\quad
                 ${}^{\ast}$ e-mail: cjf3$@$york.ac.uk
                 \\[10pt]
                 ${}^2$\,Max-Planck-Institut for
                  Mathematics in the Sciences,\\
                 Inselstr.\ 22,
                 D-04103 Leipzig, Germany\\[4pt]
                 e-mail: verch$@$mis.mpg.de\\[18pt]
\today
\end{center}
${}$\\[18pt]
{\small {\bf Abstract. } 
We study a phenomenon occuring in various areas of quantum physics, in
which an observable density (such as an energy density) which is
classically pointwise nonnegative may assume arbitrarily negative
expectation values after quantisation, even though the spatially
integrated density remains nonnegative. Two prominent examples which
have previously been studied are the energy density (in quantum field
theory) and the probability flux of rightwards-moving particles (in quantum
mechanics). However, in the quantum field context, it has been shown
that the magnitude and space-time extension of negative energy densities are not 
arbitrary, but
restricted by relations which have come to be known as `quantum
inequalities'. In the present work, we explore the extent to which
such quantum inequalities hold for typical quantum mechanical systems. 
We derive quantum inequalities of two types. The first are `kinematical'
quantum inequalities where spatially averaged densities are shown to
be bounded below. Specifically, we obtain such kinematical quantum 
inequalities for the current density in one spatial dimension (imposing 
constraints on the backflow phenomenon) and for the densities arising
in Weyl--Wigner quantization. The latter quantum inequalities are
direct consequences of sharp G{\aa}rding inequalities. The second type are
`dynamical' quantum inequalities where one obtains bounds from below
on temporally averaged densities. We derive such quantum inequalities
in the case of the energy density in general quantum mechanical systems
having suitable decay properties on the negative spectral axis of the
total energy.

Furthermore, we obtain explicit numerical values for the quantum inequalities
on the one-dimensional current density, using various spatial averaging
weight functions. We also improve the numerical value of the related
`backflow constant' previously investigated by Bracken and Melloy.
In many cases our numerical results are controlled by rigorous error estimates. 
 }
${}$

\section{Introduction}

The uncertainty principle lies at the root of many of the counterintuitive
features of quantum theory. Consider, for example, a quantum mechanical
particle moving in one dimension, whose state is a superposition of
right-moving plane waves. Although the expectation value of (any power of) its 
momentum is positive, nonetheless it is possible for the probability flux
at, say, the origin to become negative. Thus the probability of finding
the particle in the right-hand half-line can decrease!

We will return to this phenomenon, which has come to be known as {\em
backflow}~\cite{AllcockIII,BrackenMelloy,BMdirac},
in Section~\ref{sect:flux}. Another, related, phenomenon occurs in quantum
field theory. Even if one starts with a classical field theory in which energy
densities (as measured by all observers) are everywhere positive,\footnote{In 
general relativity, one would say that the field obeyed the {\em weak energy condition}
(WEC).} one finds that the renormalised energy density of the quantised field
can assume negative values~\cite{EGJ} and (in all models known to date) can even be made arbitrarily negative at
a given spacetime point by a suitable choice of state. For example, the energy 
density between Casimir plates is computed to be negative; a fact indirectly
supported by experiment (\cite{Casimir}; see the recent review \cite{Bordag.et.al}
for an exhaustive list of up to date references). Various authors have suggested employing
such effects to sustain exotic spacetime geometries containing wormholes \cite{Thorne.et.al}
or `warp drive' bubbles~\cite{Alcubierre}. Such suggestions are, however,
severely constrained~\cite{FR-worm,PF-warp} 
by the existence of bounds, known as quantum inequalities (QIs)
or quantum weak energy inequalities (QWEIs) 
\cite{AGWQI,FewsterEveson,FewsterPfenning,FTqi,FVdirac,passivity,Flanagan97,Ford78,FordRoman97,PfenningFord98}
 which impose limitations
on the magnitude and duration of negative energy densities. To give an
example, let $\langle\rho(t)\rangle_\psi$ be the energy density of the
free scalar field\footnote{Similar statements can be made for the
Maxwell, Proca  and Dirac fields~\cite{Pfenning_em,FewsterPfenning,FewsterMistry}.}
measured along an inertial worldline in Minkowski
space. Then, for any real-valued smooth compactly supported $g$, the
averaged energy density obeys~\cite{FewsterEveson,CJFTeo}
\begin{equation}
\int dt\, g(t)^2 \langle\rho(t)\rangle_\psi  \ge - \int_0^\infty du\, Q(u)|\widehat{g}(u)|^2
\end{equation}
for all physically reasonable (Hadamard) states $\psi$, where $Q$ is a
known function of polynomial growth.

The purpose of this paper is to apply techniques developed in the field
theoretic setting to quantum mechanical problems. In so doing we wish
to draw attention to a circle of ideas---including sharp G{\aa}rding inequalities,
dynamical stability and the QWEIs---which eventually ought to be seen
in the wider context of quantisation theory. 

We begin with a discussion of `kinematical QIs' in Sect.~\ref{sect:backflow}, taking 
the probability flux as our main example. We develop bounds on spatially averaged
fluxes which share some technical similarity with the QWEI proved by two of us 
for the Dirac field~\cite{FVdirac} (see also~\cite{FewsterMistry}). 
An important aspect of our treatment is the numerical
analysis of these bounds. To some extent this is motivated by the recent observation
of Marecki~\cite{Marecki} that QIs may have observational consequences in quantum optics;
it is therefore important to know how sharp analytically tractable bounds are. The
techniques used here may also be of independent interest, and we give a detailed
account in Sect.~\ref{sect:numerics}. Before that, in Sect.~\ref{sect:WignerKin}
we establish a very general form of kinematical QIs arising in the
Weyl--Wigner approach to quantizing classical systems (see, e.g., \cite{Landsman} as 
a general reference on that approach). More precisely, we consider the (quantized) configuration
space density $\mathbb{R}^n \owns x \mapsto \langle \rho_F (x)\rangle_{\psi}$ for normalized
wave-functions $\psi \in \mathscr{S}(\mathbb{R}^n)$ which are associated with classical
observables $F$, i.e.\ functions on phase space $\mathbb{R}^n \times \mathbb{R}^n$, by
\begin{equation}  
\langle \rho_F (x)\rangle_{\psi} = \int \frac{d^n p}{2 \pi} F(x,p)W_{\psi}(x,p)\,,
\end{equation}
where $W_{\psi}$ denotes the Wigner function of $\psi$. Even if $F$ is
everywhere nonnegative, the density $\langle \rho_F (x)\rangle_{\psi}$
may assume negative values owing to the indefinite sign of the Wigner
function; in fact, we show that, under very general conditions on $F$, this
quantity is unbounded above and below for arbitrary given $x$ upon varying $\psi$.
Conversely, if $F$ belongs to a certain class of symbols (in the sense of microlocal analysis
\cite{HormanderI,Taylor}) which are of second order (or lower) in the momentum variables, and if
$F$ is everywhere non-negative, then we establish a kinematical quantum inequality of the
form
\begin{equation}  
\int d^nx \, \chi(x) \langle \rho_F(x) \rangle_{\psi} \ge - C 
\end{equation}
with a suitable constant $C$ depending on the non-negative weight-function $\chi$, but not
on the (normalized) wave-function $\psi$. This is a straightforward consequence of the sharp G{\aa}rding
inequality \cite{FePh,HormanderIII,Taylor}. The general result will be illustrated by
a direct derivation of a kinematical QI for the energy density.

In Sect.~\ref{sect:dynamical} we focus attention on `dynamical' QIs which bound 
temporal averages of the energy density in 
general quantum mechanical systems. (In fact our kinematical flux inequality may be
regained as the special case, in which the evolution is the group of 
translations on the line.) These are conceptually much closer to the QIs which
have been obtained in quantum field theory; in fact, the method we use to establish
these dynamical QIs makes contact with the techniques employed in \cite{FVdirac}. Some
features of the general result will be illustrated by taking the harmonic oscillator as
a concrete example. We summarise our main results in the conclusion, Sect.~\ref{sect:conclusion}.

\section{A motivating example}\label{sect:backflow}

\subsection{Probability backflow}\label{sect:flux}

We begin with a simple example: the motion of a quantum mechanical
particle in one dimension. Some time ago, Allcock~\cite{AllcockIII}
pointed out the existence of rightwards-moving states (in which the velocity is
positive with unit probability) but for which the probability of locating
the particle in the right-hand half-line is instantaneously {\em decreasing}---a
phenomenon known as {\em probability back-flow}. This phenomenon was
subsequently studied in much greater detail by Bracken and
Melloy~\cite{BrackenMelloy} (see also~\cite{BMdirac}). To illustrate the
idea, let us suppose the normalised state $\psi$ (as well as being square-integrable itself) has
a continuous, square integrable first derivative. Then the corresponding probability
flux at position $x$ is given by
\begin{equation}
j_\psi(x)= \frac{\Re \overline{\psi(x)}(p\psi(x))}{m}\,,
\end{equation}
where the momentum operator is, as usual, $p=-i\hbar d/dx$ and the
particle has mass $m$. Now the spatial integral of the flux is
\begin{equation}
\int j_\psi(x) 
\,dx = \frac{\Re \ip{\psi}{p\psi}}{m} = \frac{\langle p\rangle_\psi}{m}
\end{equation}
and therefore yields the expected velocity. If $\psi$ is a normalised right-moving
wave-packet, it may be written by means of the Fourier transform as 
a superposition of right-moving plane waves
\begin{equation}
\psi(x) = \int \frac{dk}{2\pi} e^{ikx} \widehat{\psi}(k)
\end{equation}
with $\widehat{\psi}(k)=0$ for $k<0$, so
\begin{equation}
\langle p\rangle_\psi = \int_0^\infty\frac{dk}{2\pi}\, \hbar k
|\widehat{\psi}(k)|^2 >0
\end{equation}
and we see that the spatially integrated flux is positive. However this does not
imply that the flux itself is everywhere non-negative. Indeed, suppose that
\begin{equation}
\widehat{\psi_{k_0}}(k) = {\cal N}\chi_{[0,k_0]}(k) \left(k\sqrt{3} - k_0\right)\,,
\label{eq:bckflowex}
\end{equation}
where $\chi_{\Omega}$ denotes the characteristic function of $\Omega$
and ${\cal N}=(k_0^3(2-\sqrt{3})/(2\pi))^{-1/2}$ is a normalisation constant. One may
calculate
\begin{equation}
j_{\psi_{k_0}}(0) =
\frac{\hbar k_0^2}{4\pi m}\left(\frac{1}{2}-\frac{1}{\sqrt{3}}\right)
\sim -0.006 \frac{\hbar k_0^2}{m}\,,
\end{equation}
which is not merely negative, but 
can clearly be made as negative as we wish by tuning $k_0$. 
Because the probability flux is negative at the origin, the probability
of locating the particle in the left-half line is instantaneously
increasing, thereby providing an example of the backflow phenomenon
mentioned above. 

\begin{figure}
\psset{unit=1in}
\hfil\begin{pspicture}(0,1)(6.4,3)
\rput[l](0,2){\resizebox*{2 in}{!}{\rotatebox{270}{\includegraphics{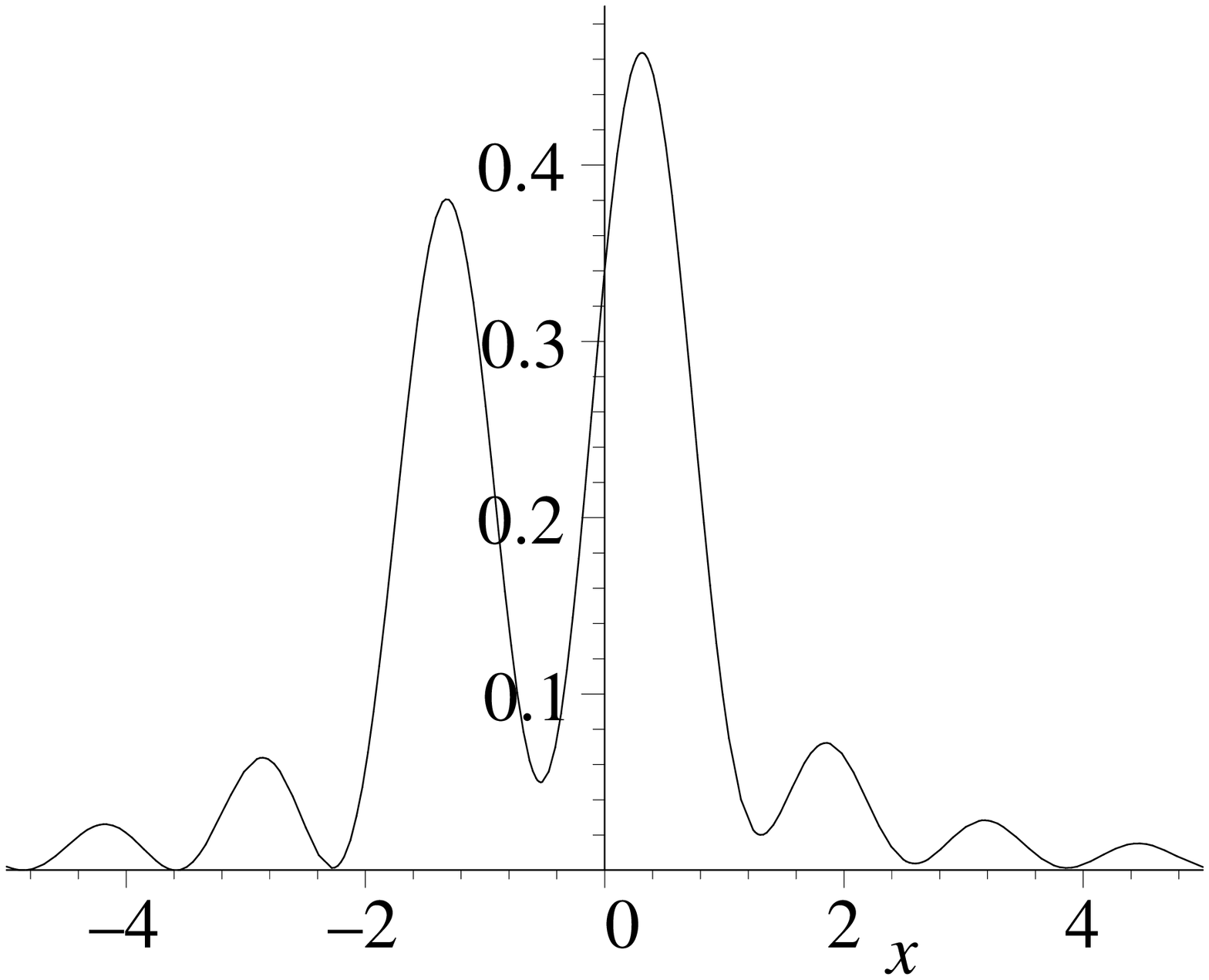}}}}
\rput[l](2.2,2){\resizebox*{2 in}{!}{\rotatebox{270}{\includegraphics{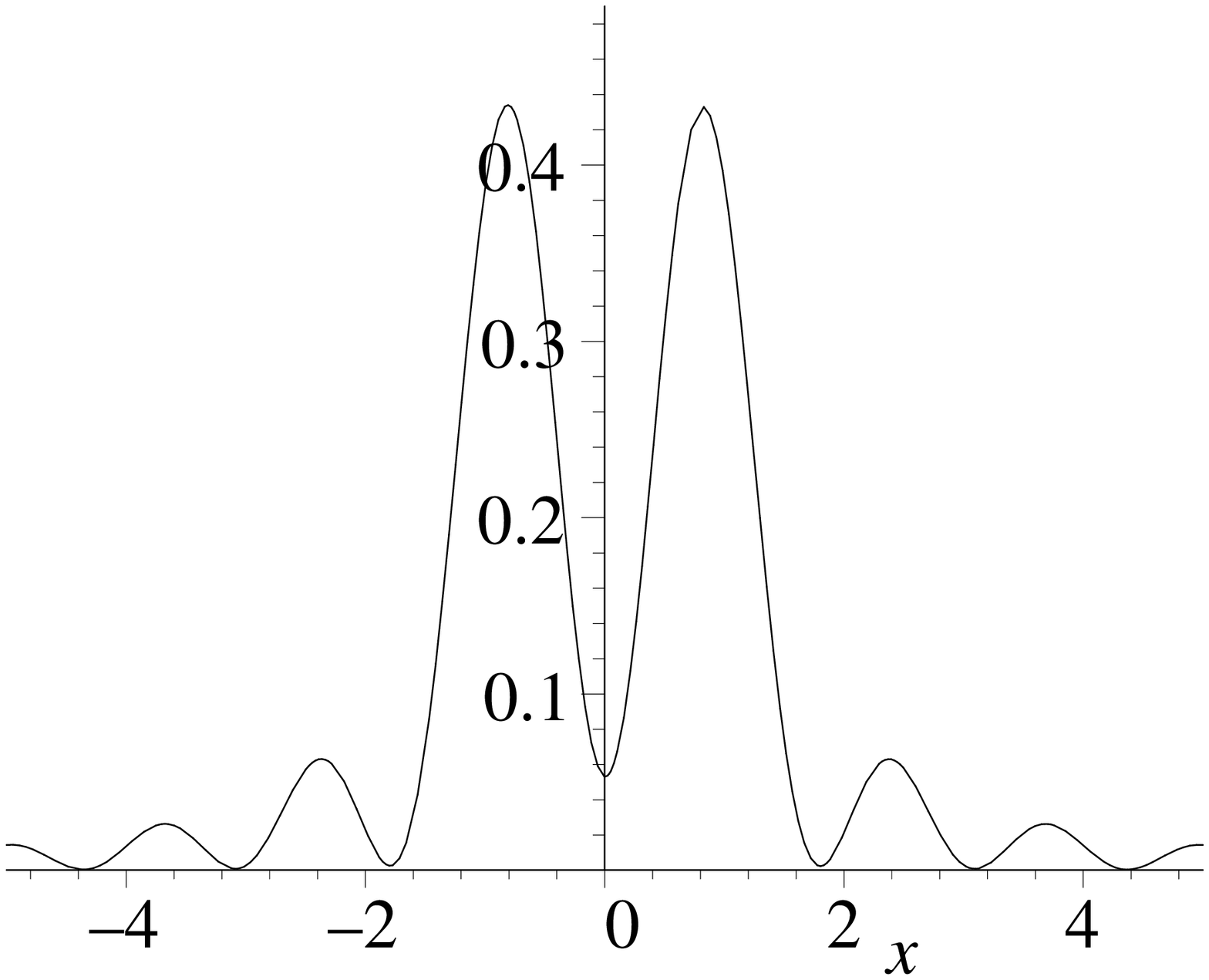}}}}
\rput[l](4.4,2){\resizebox*{2 in}{!}{\rotatebox{270}{\includegraphics{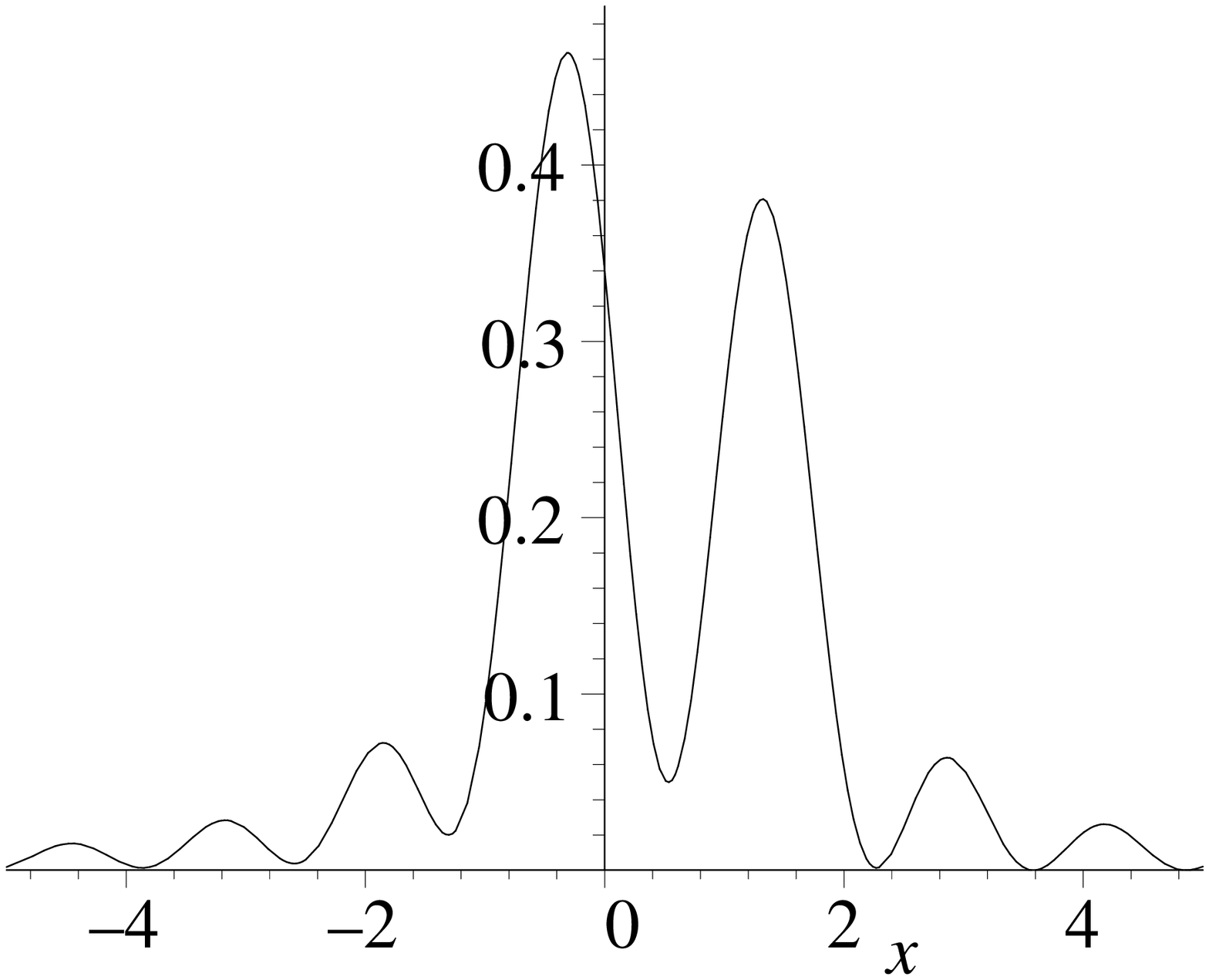}}}}
\end{pspicture}\hfil
\caption{Evolution of a wavepacket,
illustrating the backflow phenomenon. From left to right, the plots show
the position probability density at times $t=-0.1$, $t=0$ and $t=0.1$.\label{fig:backflow}} 
\end{figure}

Backflow provides a nice illustration of the inadequacy of the phase
velocity alone to predict the motion of a wavepacket. The three plots
in Figure~\ref{fig:backflow} indicate the evolution of the position
probability density in time; although the packet moves to the right, the
two main peaks are reshaped in such a way that net probability has
passed from the right-hand half line to the left. The wavepacket is
given by Eq.~\eqref{eq:bckflowex} at time $t=0$ with $k_0=5$, $m=1/2$ and $\hbar=1$.

\subsection{A quantum inequality for the flux}\label{sect:fluxQI}

\begin{figure}
\psset{unit=1in}
\hfil\begin{pspicture}(0,0)(5,4)
\rput[l](0,2){\resizebox*{5 in}{!}{\rotatebox{270}{\includegraphics{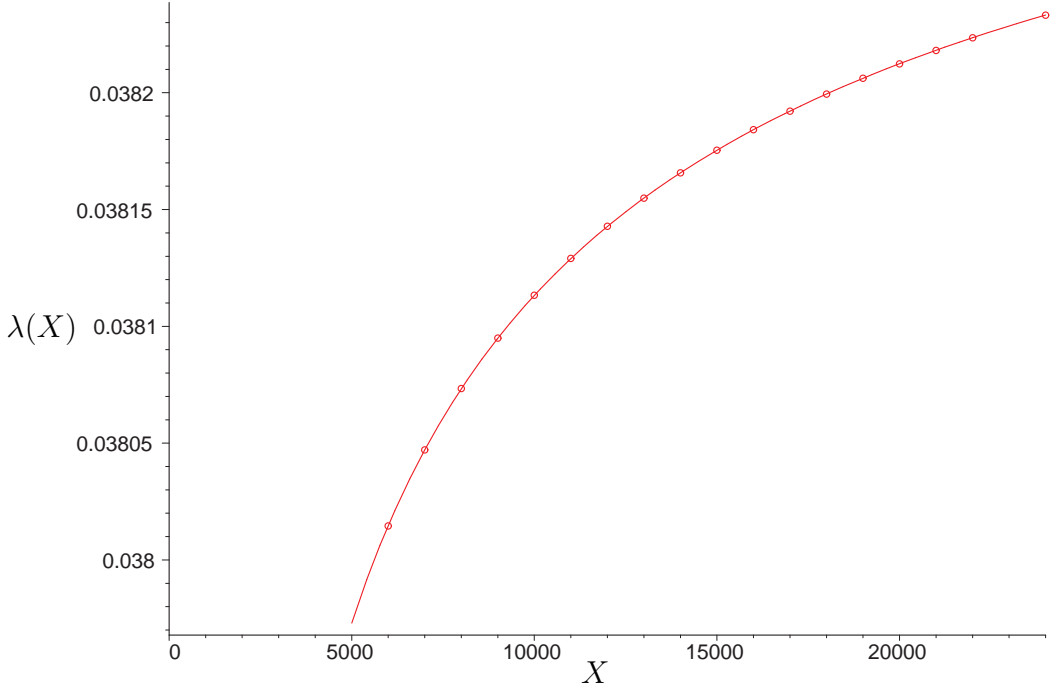}}}}
\uput[0](2.5,0.2){$X$}
\uput[180](0,2){$\lambda(X)$}
\end{pspicture}\hfil
\caption{The least squares fit of $\lambda(X)$ to $a+b/\sqrt{X}$.\label{fig:BMdata}} 
\end{figure}

As we will see in Sect.~\ref{sect:WignerKin} the backflow effect may be traced to the
uncertainty principle. From this point of view, it is natural to seek
bounds on its magnitude and extent. Bracken and Melloy~\cite{BrackenMelloy}
approached this question by showing that the probability $P(t)$ of finding
a right-moving particle in the {\em left}-hand half-line obeys
\begin{equation}
P(t) \le P(0) + \lambda\,,
\label{eq:BMbound}
\end{equation}
for all $t\ge 0$,
where the dimensionless constant $\lambda$ is the largest positive
eigenvalue of the equation
\begin{equation}
-\frac{1}{\pi}\int_0^\infty
\frac{\sin\left(u^2-v^2\right)}{u-v}\varphi(v)\,dv
=\lambda\varphi(u)
\end{equation}
(for $\varphi\in L^2(\RR^+)$). They also presented numerical evidence
that $\lambda\sim 0.04$. Using the numerical methods described in
Sect.~\ref{sect:numerics}, we have recalculated this quantity to a much higher
accuracy, although we have been unable to obtain consonant analytical
error estimates. It turns out to be convenient to change variables to $x=u^2$; we then
consider the truncation of the resulting integral kernel to $[0,X]$.
The maximum eigenvalue $\lambda(X)$ was then calculated for 
values of $X$ ranging from 6000 to 24000, using $X/2$ quadrature nodes. 
This choice was based on calculations using a variety of densities for
values of $X$ around 2000 for which $X/2$ nodes provide accuracy to 5 significant figures.
By contrast, the largest
calculation conducted in~\cite{BrackenMelloy} corresponds to $X=625$, which reflects
the increase in available computing power over the past decade.
The resulting data may be fitted  
to a remarkable degree by the form $\lambda(X)=a+b/\sqrt{X}$ (as already
noted by Bracken and Melloy for their data). Using a least squares fit to
this, we obtain the estimate $\lambda=0.03845182014$ with a maximum percentage residual
error under $4\times 10^{-4}\%$. Assuming the residual errors would be
comparable for larger $X$, this suggests that $\lambda=0.038452$ to this
level of precision. Our data points and the best-fit curve are
shown in Fig.~\ref{fig:BMdata}. 

One may interpret the Bracken--Melloy bound~\eqref{eq:BMbound} as a demonstration of the
transitory nature of backflow: large negative fluxes for right-moving
states must be short-lived. Here, we present an apparently new bound,
which demonstrates that such fluxes are also of small spatial extent, and
 whose proof is related to the
quantum weak energy inequalities derived by two of us for
the Dirac quantum field~\cite{FVdirac} (see also~\cite{FewsterMistry}). 
We consider spatially smeared quantities of the form
\begin{equation}
j_\psi(f) = \int j_\psi(x) f(x)\,dx\,,
\end{equation}
which may be regarded as the instantaneous probability flux measured by a
spatially extended detector. For any smooth, compactly supported, complex-valued
function $g$, we will show that 
\begin{equation}
\int j_{\psi}(x) |g(x)|^2 \,dx \ge -\frac{\hbar}{8\pi m}\int dx\, |g'(x)|^2
\label{eq:fluxQI}
\end{equation}
for all normalised states $\psi$ belonging to the class $\Rr$ of right-moving
states defined by 
\begin{equation}
\Rr=\{\psi\in L^2(\RR): \hbox{$\widehat{\psi}(k)=0$ for $k<0$ and
$\psi'$ continuous and square-integrable}\}\,.
\end{equation}
In fact, the conditions on both $g$ and $\psi$ may be weakened
slightly.\footnote{In particular, continuity of $\psi'$ may be weakened
to $\psi\in AC(\RR)\cap L^2(\RR)$ with $\psi'\in L^2(\RR)$ at the
expense of augmenting some statements with the qualification `almost everywhere'; by an
approximation argument it is easy to see that~\eqref{eq:fluxQI} holds
for all $g$ belonging to the Sobolev space $W^{1,2}(\RR)$~\cite{Adams}.\label{fn:generalisations}}

Before giving the proof, let us make three observations.
\begin{enumerate}
\item First, we note that there is no {\em upper} bound on the smeared flux.
To see this,
choose any normalised $\psi\in\Rr$ and let $\psi_\lambda(x)=e^{i\lambda x}\psi(x)$.
We have $\psi_\lambda\in\Rr$ for $\lambda\ge 0$; moreover, 
\begin{equation}
j_{\psi_\lambda}(x) = j_{\psi}(x) + \frac{\lambda\hbar}{m} |\psi(x)|^2
\end{equation}
so $\int j_{\psi_\lambda}(x) f(x)\,dx \to +\infty$ as
$\lambda\to+\infty$.
\item Second, the scaling behaviour of the above bound may be
investigated by replacing $g$ by
$g_\lambda(x)=\lambda^{-1/2}g(x/\lambda)$, whereupon the right-hand side
of inequality~\eqref{eq:fluxQI} scales by a factor of $\lambda^{-2}$.
The limit $\lambda\to 0$ corresponds to the unboundedness below of the
probability flux at a point, while the limit $\lambda\to\infty$ is consistent
with the fact that $\langle p\rangle_\psi\ge 0$ for $\psi\in\Rr$
(because the bound vanishes more rapidly than $\lambda^{-1}$). Roughly
speaking, our bound asserts that the magnitude of negative flux times
the square of its spatial extent satisfies a state-independent upper
bound on $\Rr$. Thus the extent of backflow is limited both in space and
in time. 

Note also that the bound~\eqref{eq:fluxQI} vanishes in both the
classical limit $\hbar\to 0$ and the limit of large mass. This differs
from Bracken and Melloy's inequality~\eqref{eq:BMbound} in which the
dimensionless constant $\lambda$ is independent of $\hbar$ and $m$. 

We remark that---again in contrast to~\cite{BrackenMelloy}---our 
result is kinematical rather than dynamical: no specific
Hamiltonian is invoked. Here, `kinematic' refers to the kinematics of
the Schr\"odinger representation, i.e., 
the (unique) regular representation of the Heisenberg commutation
relations.
\item Finally, on integration by parts, Eq.~(\ref{eq:fluxQI}) can be 
reformulated as the assertion that for each normalised $\psi\in\Rr$, the 
Schr\"odinger operator 
\begin{equation}
H_\psi= -\frac{\hbar^2}{2m}\frac{d^2}{dx^2} + 4\pi\hbar j_{\psi}(x)
\end{equation}
is positive on the space of smooth compactly supported functions $g$, 
in the sense that
\begin{equation}
\int \overline{g(x)} (H_\psi g)(x) \ge 0
\end{equation}
for all such $g$. Although the physical
significance of this reformulation is not clear, it can provide useful necessary conditions for
a given function $j(x)$ to be the flux of a right-moving state. Only if
the corresponding Schr\"odinger operator has no bound states can this be the
case. This can be sharpened slightly: as an illustration, 
suppose $j_\psi(x)$ is the flux of a state in $\Rr$ with
$j_\psi(x)\le-M$ on some open interval $I$ of length $a$. Then positivity
of $H_\psi$ on $\CoinX{I}$ implies that the Friedrichs extension $H_M$
of the operator
\begin{equation}
-\frac{\hbar^2}{2m}\frac{d^2}{dx^2} - 4\pi\hbar M \qquad {\rm
on}~\CoinX{I}\subset L^2(I)
\end{equation}
is also positive. Since the Friedrichs extension of this operator
corresponds to the imposition of Dirichlet boundary conditions at the
boundary of $I$, $H_M$ has spectrum 
\begin{equation}
E_n=\frac{\hbar^2 n^2\pi^2}{2ma^2} - 4\pi\hbar M \qquad (n=1,2,3,\ldots)
\end{equation}
so we deduce (from $E_1\ge 0$) that $M\le \hbar \pi/(8ma^2)$. This
provides a more quantitative version of the connection between the
magnitude and spatial extent of negative fluxes. 
Similar ideas have been employed in the context of quantum weak
energy inequalities~\cite{FTqi} to cast light on the `quantum interest conjecture'
of Ford and Roman~\cite{FRqi}. 
\end{enumerate}

We now establish the quantum inequality~(\ref{eq:fluxQI}). It is
sufficient to prove this for the case in which $g\in\CoinX{\RR}$ is
real-valued. Setting 
$f(x)=g(x)^2$ and writing $M_f$ for the multiplication operator
$(M_f\psi)(x) = f(x)\psi(x)$, we have
\begin{eqnarray}
\int j_{\psi}(x) f(x) \,dx &=& \frac{1}{m} \Re\ip{\psi}{M_f p \psi} \nonumber\\
&=& \frac{1}{m} \Re \left( \ip{\psi}{M_g p M_g\psi} + \ip{\psi}{M_g
[M_g,p]\psi}\right) \nonumber\\
&=& \frac{1}{m} \ip{\psi}{M_g p M_g\psi}\nonumber\\
&=& \frac{\hbar}{m} \int \frac{dk}{2\pi} k \left|\widehat{M_g\psi}(k)\right|^2\,,
\end{eqnarray}
where we have used the fact that $\Re\ip{\psi}{M_g[M_g,p]\psi} =\Re 
i\hbar\ip{\psi}{M_{gg'}\psi} = 0$. We therefore may obtain a bound by
estimating the portion of this integral arising from $k<0$:
\begin{equation}
\int j_{\psi}(x) f(x) \,dx \ge \frac{\hbar}{m} 
\int_{-\infty}^0 \frac{dk}{2\pi} k \left|\widehat{M_g\psi}(k)\right|^2
= -\frac{\hbar}{m} 
\int_0^\infty \frac{dk}{2\pi} k \left|\widehat{M_g\psi}(-k)\right|^2\,.
\label{eq:preineq}
\end{equation}
\vspace{0.1cm}

By the convolution theorem
\begin{equation}
\widehat{M_g\psi}(k) = \int_0^\infty\frac{dk'}{2\pi} \widehat{\psi}(k')
\widehat{g}(k-k')\,,
\end{equation}
where the restriction to $k'\in\RR^+$ is permissible for $\psi\in\Rr$. 
Now a straightforward application of Cauchy-Schwarz gives
\begin{equation}
\left|\widehat{M_g\psi}(-k)\right|^2 \le 
\int_0^\infty\frac{dk'}{2\pi} |\widehat{g}(k+k')|^2 \,,
\end{equation}
where we have also used $|\widehat{g}(-k)|^2 = |\widehat{g}(k)|^2$
(since $g$ is real) and $\|\psi\|=1$. Substituting in~(\ref{eq:preineq}),
we now calculate
\begin{eqnarray}
\int j_{\psi}(x) f(x) \,dx &\ge& - \frac{\hbar}{m} \int_0^\infty\frac{dk}{2\pi}
\int_0^\infty\frac{dk'}{2\pi} k |\widehat{g}(k+k')|^2\nonumber\\
&=& -\frac{\hbar}{m}\int_0^\infty \frac{du}{(2\pi)^2} |\widehat{g}(u)|^2
\int_0^u dk\, k \nonumber\\
&=& -\frac{\hbar}{m}\int_0^\infty \frac{du}{8\pi^2} u^2 |\widehat{g}(u)|^2 \nonumber\\
&=& -\frac{\hbar}{m}\int_{-\infty}^\infty \frac{du}{16\pi^2} u^2
|\widehat{g}(u)|^2 \nonumber\\
&=& -\frac{\hbar}{8\pi m}\int dx\, |g'(x)|^2\,,
\label{eq:HScalc}
\end{eqnarray}
where we have changed variables from $(k,k')$ to $(u,k)$ with $u=k+k'$,
used evenness of $|\widehat{g}(u)|$ and Parseval's theorem. This
completes the proof of the quantum
inequality~(\ref{eq:fluxQI}). 

\vspace{0.1cm}
The later stages of this argument may be rephrased as follows.
The inequality~(\ref{eq:preineq}) asserts that
\begin{equation}
\int j_{\psi}(x) f(x) \,dx \ge -\frac{\hbar}{m} \|T\widehat{\psi}\|^2
\label{eq:Tpsihat}
\end{equation}
where the operator $T$ acts on $L^2(\RR^+,dk/(2\pi))$ by
\begin{equation}
(T\varphi)(k) = \int\frac{dk'}{2\pi} \sqrt{k} \widehat{g}(-k-k') \varphi(k') 
\end{equation}
and is easily seen to be Hilbert--Schmidt. Varying over normalised
$\psi$, the right-hand side of
Eq.~(\ref{eq:Tpsihat}) is bounded below by $-\|T\|^2$, where
$\|T\|$ denotes the operator norm of $T$. This leads to the bounds
\begin{equation}
\int j_{\psi}(x) f(x) \,dx \ge -\frac{\hbar}{m} \|T\|^2 \ge -\frac{\hbar}{m}\|T\|_\hs^2\,, 
\label{eq:twobds}
\end{equation}
where the last inequality holds because the Hilbert--Schmidt norm
$\|T\|_\hs$ dominates the operator norm. The calculation
in~(\ref{eq:HScalc}) in fact precisely computes this final bound.

To summarise, we have seen that, even for a right-moving state $\psi\in\Rr$, the flux $j_\psi$ need
not be pointwise nonnegative; moreover by tuning the state, one may
arrange the normalised flux $j_\psi(x)$ to be as negative as one likes at a given
fixed $x$. However, weighted spatial averages of the flux are bounded below in
terms of the weight function alone. 
This condition may be reformulated as asserting the positivity of
the Hamiltonian for a particle moving in a potential given (up to
constants) by the probability flux of any state in $\Rr$. 

\subsection{Numerical results and sharper bounds}\label{sect:fluxnumerics}

We illustrate our bound by reference to four weight functions: a
Gaussian, a squared Lorentzian and two compactly supported weights
which we call the truncated cosine and the smoothed truncated cosine. 
(Neither of the compactly supported weights are $C^\infty$, but they have sufficient smoothness
for the above argument to hold; see footnote~\ref{fn:generalisations} above.)
Our weight functions are summarised in Table~\ref{tb:weights},
along with the corresponding bound arising from Eq.~(\ref{eq:fluxQI}).
In each case, $f_\lambda$ has unit integral, the parameter $\lambda$
controls the sampling width and $g_\lambda(x)=\sqrt{f_\lambda(x)}$.
For later reference we have also given the Fourier transforms of
$f_\lambda$ and $g_\lambda$. 

\begin{table}\rotatebox{90}{\begin{minipage}{\textheight}
\begin{tabular}{|l|l|l|l|l|}
\hline
                      &  Gaussian  & Squared Lorentzian\phantom{ZZZ} &
Truncated cosine & Smoothed truncated cosine \\ \hline\hline &&&&\\
$f_\lambda$           & $(\lambda\sqrt{\pi})^{-1} e^{-(x/\lambda)^2}$ &$2\lambda^3\pi^{-1}/(x^2+\lambda^2)^2$
& $\vartheta(\lambda-|x|)/\lambda\cos(x\pi/(2\lambda))$
& $4\vartheta(\lambda-|x|)/(3\lambda)\cos(x\pi/(2\lambda))^4$ \\[0.6cm]
$\widehat{f}_\lambda$ & $e^{-(\lambda u)^2/4}$ & $(1+\lambda |k|)e^{-\lambda|k|}$ &
$\displaystyle\frac{\pi^2\sin(\lambda k)}{\lambda k(\pi^2-k^2\lambda^2)}$
&$\displaystyle\frac{4\pi^4\sin(\lambda k)}{\lambda
k(k^2\lambda^2-4\pi^2)(k^2\lambda^2-\pi^2)}$\\[0.6cm]
$\widehat{g}_\lambda$ & $\sqrt{2\lambda}\pi^{1/4} e^{-(u\lambda)^2/2}$
& $\sqrt{2\lambda\pi}e^{-\lambda|k|}$ &
$\displaystyle\frac{4\pi\sqrt{\lambda}\cos(\lambda k)}{\pi^2-4k^2\lambda^2}$
& $\displaystyle\frac{2\pi^2\sin(\lambda k)}{k\sqrt{3\lambda}(\pi^2-k^2\lambda^2)}$ 
\\[0.6cm] 
QI bound & $-\hbar/(16\pi m\lambda^2)$ & $-\hbar/(16\pi m\lambda^2)$ & 
$-\hbar\pi/(32m\lambda^2)$
&$-\hbar\pi/(24m\lambda^2)$\\[0.2cm]
{}[$\approx\hbar/(m\lambda^2) \times$] & $-0.01989436788$ &
$-0.01989436788$ & $-0.09817477044$
 & $-0.1308996939$ \\[0.2cm]
\hline
\end{tabular}
\caption{Compendium of sampling functions considered.\label{tb:weights}}
\vspace{0.5cm}
\begin{center}\setlength{\tabcolsep}{1mm}
\begin{tabular}{|rr|rr|rr|rr|}    \hline
\multicolumn{2}{|c|}{Gaussian}  &\multicolumn{2}{c|}{Squared Lorentzian} &
\multicolumn{2}{c|}{Trunc.\ cosine}
& \multicolumn{2}{c|}{Smoothed trunc.\ cosine} \\ 
\multicolumn{1}{|c}{$K$} & \multicolumn{1}{c|}{$\mu(K)$} &
\multicolumn{1}{c}{$K$} & \multicolumn{1}{c|}{$\mu(K)$} &
\multicolumn{1}{c}{$K$} & \multicolumn{1}{c|}{$\mu(K)$} &
\multicolumn{1}{c}{$K$} & \multicolumn{1}{c|}{$\mu(K)$} \\ \hline
10 & -0.0048295212087 & 30 & -0.002980544308 &  2000  & -0.029012801924   & 140   &  -0.036095566956 \\
20 & -0.0048295668511 & 40 &                 &  2200  & -0.029012804495   & 160   &  -0.036095567038 \\ 
30 & -0.0048295668517 & 50 &                 &  2400  & -0.029012806174   & 180   &  -0.036095567056 \\ 
40 &                  & 60 &                 &  2600  & -0.029012807318   & 200   &  -0.036095567060 \\
50 &                  & 70 &                 &  2800  & -0.029012808114   & 220   &  -0.036095567061 \\
60 &                  & 80 &                 &  3000  & -0.029012808686   & 360   &  \\
\hline
\end{tabular}
\end{center}
\caption{Numerical estimation of $\inf\sigma(J)$ for various kernels.\label{tb:unbounded}}
\end{minipage}}
\end{table}

We wish to compare the above bound with two sharper (but less
analytically tractable) bounds: the bound arising from the first inequality
in~(\ref{eq:twobds}) and a direct numerical estimate of the infimum of 
the integrated flux. In the first case, we are required to find the
operator norm of $T$. Our numerical approach proceeds by first truncating
the kernel to an interval $[0,K]$---we are able to estimate the
error incurred here by using bounds obtained from the Hilbert--Schmidt
norm---and applying a numerical quadrature scheme due originally to Fredholm (see
e.g., Sec.~4.1 of~\cite{Atkinson} or Chapter~4 of~\cite{DelvesMohamed}) to the truncated
kernel. This leads to a matrix whose eigenvalues approximate those
of the truncated kernel and hence the original operator, and which can
be computed using standard numerical packages. Full details, 
including a discussion of error estimates, are given in Sect.~\ref{sect:numerics}. This leads
to quantum inequalities
\begin{equation}
\int j_{\psi}(x) f_\lambda(x) \,dx
\ge -\frac{\hbar C}{m\lambda^2}\,,
\end{equation}
where the constant $C$ depends on the particular weight function used
and is given:
\begin{center}
\begin{tabular}{|c|l|l|l|}
\hline
Kernel & $C$ & Accuracy & Improvement \\ \hline\hline
Gaussian & 0.01958128485 & 10 S.F. & 1.6\% \\
Squared Lorentzian & $(16\pi)^{-1}$ & Exact & 0 \\
Trunc. cosine & 0.08463957004 & 2 S.F.? & 16\% \\
Smooth trunc. cosine  & 0.125047838 & At least 3 S.F. & 4.5\% \\
\hline
\end{tabular}
\end{center}
Note that we were only able to obtain fairly weak error bounds in the
truncated cosine case. 
In each case, the improvement on the analytical bound is relatively
small. We may interpret these
results as showing that $T=R+S$ where $R$ has rank $1$ and the
Hilbert--Schmidt norm of $S$ is small relative to that of $T$. This
is most apparent in the squared Lorentzian case, in which $T$ is itself
exactly rank $1$ and no improvement is obtained by using the operator
norm. It would be interesting to understand the origin of this
apparently general phenomenon. 

Our second numerical calculation aims to compute the infimum of the spectrum
of the unbounded integral operator
\begin{equation}
(J\varphi)(k)= \frac{\hbar}{2}\int_0^\infty\frac{dk'}{2\pi}\,
(k+k')\widehat{f}_\lambda(k-k') \varphi(k')\,.
\label{eq:Jdef}
\end{equation}
We proceed by truncating the kernel to the interval $[0,K]$, computing
the minimum eigenvalue $\mu(K)$ using sufficiently many quadrature
points to obtain machine precision. We then increase $K$ until
convergence of $\mu(K)$ is obtained, again to machine precision. Our
results are given in Table~\ref{tb:unbounded}, in which we give
$\mu(K)$ in units of $\hbar/(m\lambda^2)$. Blank entries indicate
that the computed value was identical to the last printed number in that
column. The density of quadrature points used (per unit $K$) was 5 for
the Gaussian, 1 for the truncated cosine, and 5 for the smoothed
truncated cosine, although higher densities were also used as a
numerical check (40, 2, and 10 respectively). The results for the 
squared Lorentzian were rather slower to converge as the density
increased (perhaps because the kernel fails to be everywhere smooth) and
were computed using a density of 60. For $K<80$, a density of 70 was
used as a check.

To summarise, we have seen that a) the limitations of our flux QI do not
lie in the estimation of an operator norm by a Hilbert--Schmidt norm,
but rather in the earlier stages of the derivation (probably the
estimate~\eqref{eq:preineq}); b) the overall scope for improvement on
our flux QI is roughly a factor of between 3 and 7 (in our examples),
and it is clear that the sharp bound is not simply a multiple of our
bound~\eqref{eq:fluxQI} (in contrast to the situation for
two-dimensional massless quantum fields~\cite{Flanagan97,FewsterEveson}).

\section{The Wigner function and Kinematical Quantum Inequalities}\label{sect:WignerKin}

It is worth emphasising that phenomena similar to those presented above
arise naturally in the context of Weyl quantisation, in which the
phase space aspect of quantum mechanics is brought to the fore. In our
discussion we will consider the phase space to be $\RR^n\times\RR^n$
(see, e.g.,~\cite{Landsman} for Weyl quantisation on manifolds). 
We recall that the central object in this approach is the Wigner
function $W_\psi$ defined on phase space by
\begin{equation}
W_\psi(x,p)=\left(\frac{2}{\hbar}\right)^n
\int d^ny\, e^{2ipy/\hbar} \overline{\psi(x+y)}\psi(x-y)\,,
\end{equation}
where $\psi\in L^2(\RR^n)$ is the corresponding normalised quantum mechanical state
vector. The classical analogue of $W_\psi(x,p)$ would be a probability
distribution on phase space; as is well known, however, $W_\psi$ is not
itself a probability distribution because it is not guaranteed to be
everywhere nonnegative. This has important consequences for observables
obtained via Weyl quantisation, which proceeds as follows.

Given an observable on the classical phase space, i.e., 
a smooth function\footnote{Precise growth conditions will be specified
below.} $F:\RR^{n}\times\RR^n\to\RR$, Weyl quantisation
defines an operator $F_w$ whose expectation values are given (for
normalised $\psi$) by
\begin{equation}
\langle F_w\rangle_\psi = \int \frac{d^nx\,d^np}{(2\pi)^n} F(x,p) W_\psi(x,p)\,.
\label{eq:Fwexp}
\end{equation}
The action of this operator may be written in the form 
\begin{equation}
(F_w\psi)(x) = \int \frac{d^ny\,d^np}{(2\pi\hbar)^n} F([x+y]/2,p)
e^{i(x- y)\cdot p/\hbar} \psi(y) \,.
\end{equation}

Let us note that this procedure also yields a natural definition for the
quantum mechanical density associated with a classical observable.
Namely, setting
\begin{equation}
\langle\rho_F(x)\rangle_\psi = \int \frac{d^np}{(2\pi)^n} F(x,p) W_\psi(x,p)\,,
\label{eq:densities}
\end{equation}
it is clear that the spatial integral of $\langle\rho_F(x)\rangle_\psi$ 
yields the expectation value $\langle F_w\rangle_\psi$ for all
$F$ and $\psi$ (modulo domain questions\footnote{For example, this will
certainly hold for $F$ of polynomially bounded growth and $\psi$
belonging to the Schwartz class.}). 

Now, because the Wigner function need not be everywhere positive,
we see that the Weyl quantisation of a nonnegative
classical observable may assume negative expectation values. This
situation is exacerbated for the densities defined above (see statement
(II) below). However, we
will show that kinematical quantum inequalities may be derived, under
certain conditions. Indeed, these bounds are obtained as applications of
the so-called sharp G{\aa}rding inequalities in the theory of
pseudodifferential operators \cite{Taylor,FePh,HormanderIII}.
It is interesting to note that Fefferman and Phong, to whom the most
general sharp G{\aa}rding results are due, were guided by intuition
arising from quantum mechanics: in particular, the uncertainty
principle.

We begin by specifying more precisely the class of classical
observables. For $m\in\NN$, 
the symbol class $S^m$ (often denoted more precisely as $S^m_{1,0}$)
is defined to be the set of smooth functions $F:\RR^n\times\RR^n\to\CC$
such that, for each compact $K\subset\RR^n$ and $n$-dimensional
multi-indices 
$\alpha,\beta$, there exists a constant $C_{K,\alpha,\beta}$ such that
\begin{equation}
\label{eq:Sm}
|(D_x^\alpha D_p^\beta F)(x,p)| \le C_{K,\alpha,\beta}
(1+|p|)^{m-|\beta|}
\end{equation}
for all $(x,p)\in K\times \RR^n$
(see, e.g., \cite{HormanderI,Taylor} for multi-index notation). By $S^m_{\sf
hom}$, we denote the set of $F\in S^m$ admitting a (unique) decomposition 
$F=F_{\sf pr} + F_{\sf sub}$ such that the 
principal symbol $F_{\sf pr}$ belongs to $S^m$, is homogeneous of
degree $m$ in momentum, i.e., $F_{\sf pr}(x,\lambda p)=\lambda^m F_{\sf
pr}(x,p)$ for all $(x,p)\in\RR^n$ and $\lambda\in\RR$, and is 
nonzero except at vanishing momentum,
while the sub-principal symbol $F_{\sf sub}$ belongs to $S^{m-1}$. 

For $F\in S^m$, the Weyl quantisation $F_w$ is a continuous linear map 
from $\CoinX{\RR^n}$ to $C^\infty(\RR^n)$, so Eq.~\eqref{eq:Fwexp} holds
for all normalised $\psi\in\CoinX{\RR^n}$. The density
$\langle\rho_F(x)\rangle_\psi$ is in fact defined (and indeed smooth in
$x$) for all $\psi$ belonging to the Schwartz class $\Sch(\RR^n)$;
however, it is only guaranteed to be integrable for
$\psi\in\CoinX{\RR^n}$. 

Below, we will establish the following observations:
\begin{itemize}
\item[(I)] Suppose $F\in S^m_{\sf hom}$ is real, for some $m\ge 1$. 
Then, for each $x$, 
the density $\langle \rho_F(x)\rangle_\psi$ is unbounded from both above and below
as $\psi$ varies in $\CoinX{\RR^n}$ with $||\psi ||_{L^2} = 1$.
\item[(II)] Suppose $F\in S^2$ is nonnegative, $F(x,p)\ge 0$ for all
$(x,p)$ and let $\chi\in\CoinX{\RR^n}$ be nonnegative. Then there exists
a constant $C\ge 0$, depending on $F$ and $\chi$, such that
\begin{equation}
\int d^nx\, \chi(x) \langle \rho_F(x)\rangle_\psi \ge -C
\end{equation}
for all $\psi\in\Sch(\RR^n)$ with $||\psi ||_{L^2} = 1$. 
\end{itemize}

To establish (I) we may assume without loss of generality that $x=0$,
for which we have
\begin{equation}
\langle \rho_F(0)\rangle_\psi = \left(\frac{2}{\hbar}\right)^n\int \frac{d^np\,
d^ny}{(2\pi)^n}\, F(0,p) e^{2ipy/\hbar}
\overline{\psi(y)}\psi(-y)
\end{equation}
for normalised $\psi\in\CoinX{\RR^n}$. Setting $\psi_\lambda(x) =
\lambda^{-n/2}\psi(x/\lambda)$, ($\lambda>0$) and making the obvious
change of variables,
\begin{equation}
\langle \rho_F(0)\rangle_{\psi_\lambda} = \left(\frac{2}{\lambda\hbar}\right)^{n}
\int \frac{d^np\, d^ny}{(2\pi)^n}\, F(0,p/\lambda) e^{2ipy/\hbar}
\overline{\psi(y)}\psi(-y)
\end{equation}
so, bearing in mind that
\begin{equation}
|F(0,p/\lambda)-F_{\sf pr}(0,p/\lambda)|\le C(1+|p/\lambda|)^{m-1}
\end{equation}
by definition of $S^m_{\sf hom}$ and Eq.~\eqref{eq:Sm}, we obtain
\begin{equation}
\lambda^{m+n}\langle \rho_F(0)\rangle_{\psi_\lambda}\longrightarrow
\left(\frac{2}{\hbar}\right)^n\int \frac{d^np\, d^ny}{(2\pi)^n}\, F_{\sf pr}(0,p) e^{2ipy/\hbar}
\overline{\psi(y)}\psi(-y)
\label{eq:FprW} 
\end{equation}
as $\lambda\to 0^+$. It now remains to show that the right-hand side of
this expression attains values of both signs as $\psi$ varies in $\CoinX{\RR^n}$.
To this end, assume (without loss) that $F_{\sf pr}(0,p)$ depends
nontrivially on the first coordinate, $p_1$, of $p$. Integrating by parts, the right-hand side
of~\eqref{eq:FprW} in the form $P_y
(\overline{\psi(y)}\psi(-y))|_{y=0}$, where $P_y$ is a 
homogeneous linear partial differential operator (in $y$) of order $m$ with 
(possibly complex) constant coefficients $c_\alpha$. We now consider
$\psi$ of the form
\begin{equation}
\psi(y) = f(y_1) e^{i(y_2+\cdots+y_n)}\chi(y)
\end{equation}
where $\chi\in\CoinX{\RR^n}$ is equal to unity in a neighbourhood of the origin
and $f\in\CoinX{\RR}$. For such $\psi$ we have
\begin{equation}
P_y (\overline{\psi(y)}\psi(-y))|_{y=0} = Q_{y_1} 
(\overline{f(y_1)}f(-y_1))|_{y_1=0}
\end{equation}
for some ordinary differential operator
\begin{equation}
Q_{y_1} = \sum_{k=0}^q c_k (-i)^k \frac{d^k}{dy_1^k}
\end{equation}
of order $1\le q\le m$ with constant real coefficients. (That $Q_{y_1}$
is of order at least one is a consequence of our assumption that $F_{\sf
pr}(0,p)$ depends nontrivially on $p_1$; reality of the $c_k$ holds
because the right-hand side of Eq.~\eqref{eq:FprW} is manifestly real
for all $\psi\in\CoinX{\RR^n}$.) We now choose $f$ so that $f(0)=1$ and
$f^{(k)}(0)=0$ for $1\le k\le q-1$. Then by Leibniz' formula,
\begin{equation}
P_y (\overline{\psi(y)}\psi(-y))|_{y=0} = c_q
(-i)^q\left(\overline{f^{(q)}(0)}
+ (-1)^q f^{(q)}(0)\right) + c_0\,.
\end{equation}
It is now obvious that $f$ may be chosen so that the right-hand side of
this expression adopts values of both signs, completing the argument.

Statement (II) is straightforward: because $\chi\in\CoinX{\RR^n}$ and
$F\in S^2$, the symbol $\chi F$ obeys uniform bounds
\begin{equation}
|(D_x^\alpha D_p^\beta \chi F)(x,p)| \le C_{\alpha,\beta}
(1+|p|)^{m-|\beta|}
\end{equation}
for all $(x,p)\in \RR^n\times \RR^n$, so the
sharp G{\aa}rding inequality (Corollary~18.6.11 in~\cite{HormanderIII}
with $\delta=0$, $\rho=1$; see also Eq.~(18.1.1)${}''$ therein) 
entails the existence of a constant $C$ such that
\begin{equation}
\int d^n x\, 
 \chi(x) \langle \rho_F(x)\rangle_\psi =
\langle (\chi F)_w\rangle_\psi \ge -C 
\end{equation}
for all normalised $\psi\in\Sch(\mathbb{R}^n)$. This is the required kinematical quantum inequality.
\\[6pt]
We now give two examples to illustrate the above ideas. \\
${}$ \\
{\noindent\em Example 1:} Consider a
classical Hamiltonian
\begin{equation}
H(x,p) = \frac{p^2}{2m} + V(x)
\end{equation}
on $\RR^n\times\RR^n$ with $V\in C^\infty(\RR^n)$. The Hamiltonian density obtained from
Eq.~\eqref{eq:densities} is
\begin{equation}
\langle \rho_H(x) \rangle_\psi =  
\frac{\hbar^2}{4m}\left(|\nabla\psi(x)|^2-\Re\overline{\psi(x)}(\triangle\psi)(x)\right)
+V(x)|\psi(x)|^2\,,
\end{equation}
where $\triangle=\nabla^2$ is the Laplacian. Clearly $\langle \rho_H(x)
\rangle_\psi$ may be made arbitrarily negative as $\psi$ varies in
$\CoinX{\RR^n}$ by arranging that
$\nabla\psi(x)=0$, $\overline{\psi(x)}\triangle\psi(x)>0$ and then---as
in (I) above---scaling $\psi$ about $x$, introducing
\begin{equation}
\psi_\lambda(y) = \lambda^{-n/2}\psi((y-x)/\lambda)
\end{equation}
for which
\begin{equation}
\langle \rho_H(x)\rangle_{\psi_\lambda} = 
-\lambda^{-(n+2)}\frac{\hbar^2}{4m}\Re\overline{\psi(x)}(\triangle\psi)(x)
+\lambda^{-n}V(x)|\psi(x)|^2\,.
\end{equation}
As in the proof of (I), the subprincipal symbol drops out in the limit
$\lambda\to 0^+$, so $\langle\rho_H(x)\rangle_{\psi_\lambda}\to-\infty$

Since $H\in S^2$ we already know that a kinematical quantum inequality
exists. However it is instructive to give a direct argument for this,
which also yields an explicit bound. To this end, we note that, for any
nonnegative $\chi\in\CoinX{\RR^n}$ and normalised $\psi\in\Sch(\RR^n)$,
\begin{equation}
\int d^nx\,\chi(x)\langle \rho_H(x)\rangle_\psi =
\frac{1}{4m}\ip{p_i\psi}{M_\chi p_i\psi}+
\frac{1}{8m}\ip{\psi}{(M_\chi p^2+p^2M_\chi)\psi}+\ip{\psi}{M_{\chi V}\psi}
\end{equation}
where $p_i=-i\hbar\nabla_i$ and $(M_f\psi)(x)=f(x)\psi(x)$ is the
operator of multiplication by $f$. Now $[M_\chi,p]=i\hbar M_{\nabla_i \chi}$, so
\begin{equation}
M_\chi p^2+p^2M_\chi = 2p_iM_\chi p_i + i\hbar[M_{\nabla_i\chi},p_i]
=2p_i M_\chi p_i -\hbar^2 M_{\triangle\chi}
\end{equation}
and hence
\begin{equation}
\int d^nx\,\chi(x)\langle \rho_H(x)\rangle_\psi =
\frac{1}{2m}\ip{p_i\psi}{M_\chi p_i\psi}+\ip{\psi}{M_L\psi}
\label{eq:pre_kin_QI}
\end{equation}
where
\begin{equation}
L(x) = -\frac{\hbar^2}{8m} (\triangle \chi)(x)+ V(x)\chi(x)\,.
\end{equation}
Since the first term in~\eqref{eq:pre_kin_QI} is nonnegative, we obtain
the quantum inequality
\begin{equation}
\int d^nx\,\chi(x)\langle \rho_H(x)\rangle_\psi \ge 
\inf_{x\in\RR^n} \left(
-\frac{\hbar^2}{8m} (\triangle \chi)(x)+ V(x)\chi(x)\right)\,,
\end{equation}
for all nonnegative $\chi\in\CoinX{\RR^n}$ and $\psi\in \Sch(\RR^n)$. We note
as a curiosity the appearance of a Schr\"odinger operator applied to the weight $\chi$
(rather than the state $\psi$). In the case of a nonnegative potential,
we may obtain a QI (slightly weaker than that given above) in the form
\begin{equation}
\int d^nx\,\chi(x)\langle \rho_H(x)\rangle_\psi \ge \frac{1}{4}\inf_{x\in\RR^n}
(H\chi)(x) \,.
\end{equation}
${}$ \\
{\noindent\em Example 2:} We now show that (II) allows us to deduce the
existence of a kinematical flux QI on rightwards moving states. Let $f$
be a nonnegative smooth compactly supported function.   
The averaged probability flux $j_\psi(f)$ is easily seen to be the
expectation in state $\psi$ of the Weyl quantisation $j(f)$ of $f(x)p/m$,
which is a (first order) 
element of the symbol class $S^2$ (but is of course negative for $p<0$). Now let
$\eta(p)$ be smooth, vanishing for $p<0$ and equal to $p$ for $p$
greater than some $p_0$, and set $F(x,p)=f(x)\eta(p)/m$. Then the
quantisation $F_w$ differs from $j(f)$ on $\Rr\cap\Sch(\RR)$
only by a bounded operator. Accordingly (II) entails that $j_\psi(f)$ 
is bounded below for normalised $\psi\in\Rr\cap\Sch(\RR)$.  Of
course, this argument does not determine the magnitude of the bound, in
contrast to the direct approach of Sect.~\ref{sect:fluxQI}.\\
${}$ \\
We should like to remark that in \cite{BrackenDoebnerWood} there appears a
result which is complementary to ours; in that reference, the authors consider
the one-dimensional case $n=1$ and show that there is a 
$\psi$-independent bound below (and above) on the
integral of the Wigner function over elliptic sub-regions of the phase-plane
which is much sharper than that implied by the a priori 
uniform bounds
on the Wigner function. This is again an effect of averaging, this
time over a region of finite extension in both $x$- and $p$-space.
It would be interesting to see if this result can be generalized to
higher dimensions through a generalization of (II) to a more general class of
symbols; however, it is not at all clear that this can be accomplished 
as it apparently goes beyond the scope of sharp G{\aa}rding inequalities.

\section{Dynamical quantum inequalities}\label{sect:dynamical}

In this section we turn to a different type of QI, which is closer to
those studied in quantum field theory. The focus here is on
time-averages of the energy density at a fixed spatial point: we will
refer to the QI bounds obtained as dynamical quantum inequalities. To keep the discussion fairly
general, we assume that configuration
space $M$ is a topological space carrying a measure $\nu$, so that the
state space is $\HH=L^2(M,d\nu)$. The dynamics is assumed to be generated
by a self-adjoint Hamiltonian $H$
which is defined on a dense domain in $\HH$. Each normalised state
$\psi$ belonging to the domain of $H$ then determines both a position
probability density $\langle\rho(t,x)\rangle_\psi$ and a Hamiltonian
density $\langle h(t,x)\rangle_\psi$ by
\begin{eqnarray}
\langle\rho(t,x)\rangle_\psi &=& |\psi_t(x)|^2 \\ 
\langle h(t,x)\rangle_\psi &=& \Re \overline{\psi_t(x)}(H\psi_t)(x)
\end{eqnarray}
where $\psi_t=e^{-iHt/\hbar}\psi$. This definition of the energy density
differs from that employed in Sect.~\ref{sect:WignerKin}; note that we are not
assuming in this section that $H$ is the quantisation of a classical
observable, in which case the above would appear to be the most natural
definition. In particular, both the quantities defined are integrable
with respect to the measure $d\nu(x)$ for each $t\in\RR$, with integrals
equal to unity and $\langle H\rangle_\psi$ respectively. However, we 
will be interested mainly in time averages of these quantities at some fixed point $x\in M$.
In so doing, we immediately encounter the problem that it does not
generally make sense to speak of the value of an $L^1$-function at a
point.\footnote{Elements of the space $L^1(M,d\mu)$ are really
equivalence classes of functions agreeing almost everywhere.} To avoid
this, we introduce the spaces $\HH_k=D((1+H^2)^{k/2})$ and assume that,
for some $k>0$, each element in $\HH_k$ should be
(almost everywhere equal to) a continuous function and that for each
$x\in M$ there is a vector $\eta_x\in\HH$ such that
\begin{equation}
\psi(x) = \ip{\eta_x}{p_k(H)\psi} \qquad\forall \psi\in \HH_k\,.
\label{eq:eta_delta}
\end{equation}
Here, $\psi(x)$ means the value at $x$ of the continuous function
to which $\psi$ is almost everywhere equal, and we have written
$p_k(E)=(1+E^2)^{k/2}$. Therefore, the functional
$\psi\mapsto \ip{\eta_x}{p_k(H)\psi}$ on $\HH_k$ coincides with the
$\delta$-distribution concentrated at $x$, so that formally [as it is
not an element of $\HH$] $p_k(H)^*\eta_x$ is 
the $\delta$-distribution. In practice, these assumptions are fairly
mild: in particular, for the case in which $H$ is minus the
Laplacian on some manifold they are simply a transcription of the
content of Sobolev's lemma. We remark that the necessary regularity in
quantum field theoretic quantum inequalities is obtained by restricting
to the class of Hadamard states, which would correspond to 
$\HH_{\infty}=\bigcap_{k\in\NN} \HH_k$ in the present context.

It now makes sense to define the position and Hamiltonian densities as
\begin{eqnarray}
\langle\rho(t,x)\rangle_\psi &=& |\ip{\eta_x}{p(H)\psi_t}|^2 \\ 
\langle h(t,x)\rangle_\psi &=& \Re \ip{p(H)\psi_t}{\eta_x}\ip{\eta_x}{p_k(H)H\psi_t}
\end{eqnarray}
for normalised states $\psi\in\HH_{k+1}$.
Furthermore, one may easily check
(using Cauchy-Schwarz) that these quantities are bounded in $t$, so the
time-averaged quantities $\langle \rho_x(f)\rangle_\psi$ and $\langle
h_x(f)\rangle_\psi$ given by
\begin{equation}
\langle \rho_x(f)\rangle_\psi = \int dt\, f(t)
\langle\rho(t,x)\rangle_\psi
\end{equation}
and the analogous equation for $\langle h_x(f)\rangle_\psi$ are
well-defined for any smooth compactly supported function $f$. 

{}From now on, we denote the spectral measure of $H$ by $dP_E$. 
(In the case where $H$ may be diagonalised by a basis of orthogonal
eigenvectors $\phi_n$ with simple eigenvalues $E_n$, 
\begin{eqnarray}
\int d\ip{\psi}{P_E\varphi} f(E) &=& \sum_{n} 
\ip{\psi}{\phi_n}\ip{\phi_n}{\varphi} f(E_n) \nonumber\\
&=& \int dE \sum_n \delta(E-E_n)
\ip{\psi}{\phi_n}\ip{\phi_n}{\varphi} f(E);
\end{eqnarray}
more generally, the projection-valued measure allows for the case of
varying--even infinite--multiplicities and for both continuous and
discrete spectrum.) While there is some ambiguity in choosing $k$ and $\eta_x$ such that
Eq.~\eqref{eq:eta_delta} holds, the measure on $\RR$ defined by
\begin{equation}
\mu_x(\Delta) = \int_\Delta \ip{\eta_x}{dP_E\eta_x}\,p_k(E)^2
\end{equation}
for bounded Borel sets $\Delta$ has an independent meaning. In fact,
$\mu_x(\Delta)$ is simply the diagonal $P_\Delta(x,x)$ of the integral kernel
of the spectral projection $P_\Delta$ of $H$ on $\Delta$,\footnote{Since, for any
$\psi\in\HH$,  we have $P_\Delta\psi\in D(p(H))$, it follows that
$(P_\Delta\psi)(x) = \int d\nu(y)\, P_\Delta(x,y)\psi(y)$ where 
$P_\Delta(x,y) = \overline{(p_k(H)P_\Delta\eta_x)(y)}$ is continuous in
$y$. This last quantity may easily be expressed as
$\ip{p(H)P_\Delta\eta_x}{p_k(H)P_\Delta\eta_y}$, so in particular, 
$P_\Delta(x,x) = \|p_k(H)P_\Delta\eta_x\|^2 = \mu_x(\Delta)$.}
given by 
\begin{equation}
\mu_x(\Delta) = \sum_{n: E_n\in\Delta} |\phi_n(x)|^2\,,
\end{equation}
if $H$ has purely discrete spectrum.  
Below, it will occasionally be useful to consider the corresponding
measure arising from self-adjoint operators other than $H$; in these
cases, we will write $\mu_x^{(H')}$ to denote the operator $H'$ involved. 
Finally, since $0\le \mu_x(\Delta)\le \|\eta_x\|^2\sup_{E\in\Delta}
p_k(E)^2$, we see that $\mu_x$ is polynomially bounded.

After these preliminaries, we come to the statement of our dynamical
quantum inequalities. 

\begin{itemize}
\item[(III)] Let $g$ be any real-valued, compactly supported function on
$\RR$ and set $f=g^2$. Then given real numbers $a<b$, the inequalities
\begin{equation}
b\langle\rho_x(f)\rangle_\psi + \int \frac{du}{2\pi}\,Q_+(u)|\widehat{g}(u)|^2
\ge \langle h_x(f)\rangle_\psi
\ge a\langle\rho_x(f)\rangle_\psi - \int
\frac{du}{2\pi}\,Q_-(u)|\widehat{g}(u)|^2
\label{eq:III}
\end{equation}
hold for all normalised $\psi\in P_{[a,b]}\HH$, where
\begin{eqnarray}
Q_-(u) &=& \int_{[a,b]} d\mu_x(E) \{\hbar u+a-E\}_+ \nonumber\\
Q_+(u) &=& \int_{[a,b]} d\mu_x(E) \{\hbar u-b+E\}_+
\label{eq:Qsdef}
\end{eqnarray}
are nonnegative, monotone increasing and polynomially bounded in $u$,
and we have used the notation $\{\lambda\}_+=\max\{0,\lambda\}$.
(Similarly, we will write $\{\lambda\}_-=\min\{0,\lambda\}$.)
Moreover, the first (resp., second) inequality in~\eqref{eq:III} also
holds for all $\psi\in P_{(-\infty,b]}\HH_{k+1}$ (resp.,
$P_{[a,\infty)}\HH_{k+1}$) provided the integration range
in~\eqref{eq:Qsdef} is replaced by $(-\infty,b]$ (resp., $[a,\infty)$).

\item[(IV)] Suppose $\int_{\RR^-} d\mu_x(E) (1+|E|)<\infty$ and let $g$
be as in (III). Then, for
any fixed $c\in\RR$, the inequality
\begin{equation}
\langle h_x(f)\rangle_\psi
\ge c\langle\rho_x(f)\rangle_\psi - \int \frac{du}{2\pi}\, S(H-c\II;u) |\widehat{g}(u)|^2
\label{eq:IVstmnt}
\end{equation}
holds for all $\psi\in\HH_{k+1}$, where
\begin{equation}
S(H;u) = \int d\mu^{(H)}_x(E)\, \{\hbar u-E\}_+
\end{equation}
is nonnegative, monotone increasing and polynomially bounded in $u$.
(There is of course a dual statement, for the case $\int_{\RR^+} d\mu_x(E)
(1+|E|)<\infty$.)
\end{itemize}

Before proceeding to the proof of these statements, we illustrate them
by drawing some consequences. 
The interpretation of (III) is that a state with energy
between $a$ and $b$ has an averaged energy density between $a$ and $b$,
suitably weighted by the averaged position probability density, modulo a
certain latitude bounded by quantum inequalities. Replacing $g$ by
$g_\lambda(t)=\lambda^{-1/2}g(t/\lambda)$, we may consider the two
regimes $\lambda\to 0^+$, representing tightly peaked averages, and
$\lambda\to\infty$, which represents widely spread averages. In the
former case, we have
\begin{equation}
\int\frac{du}{2\pi}\, Q_\pm(u) |\widehat{g}_\lambda(u)|^2 \sim
\frac{\hbar\mu_x([a,b])}{\lambda}\int_0^\infty \frac{du}{2\pi}\, u 
|\widehat{g}(u)|^2 
\end{equation}
provided $\mu_x([a,b])>0$ (failing which the left-hand side vanishes
identically). Thus the latitude afforded by the  quantum inequality
bound grows as the sampling becomes more tightly peaked. As
$\lambda\to\infty$, the QI latitude tends to zero and one may show that
\begin{equation}
b\ge \limsup_{\lambda\to\infty}
\frac{\langle h_x(g_\lambda^2)\rangle_\psi}{\langle \rho_x(g_\lambda^2)\rangle_\psi}
\ge \liminf_{\lambda\to\infty}
\frac{\langle h_x(g_\lambda^2)\rangle_\psi}{\langle \rho_x(g_\lambda^2)\rangle_\psi}
\ge a
\end{equation}
for all $\psi\in P_{[a,b]}\HH$, provided $\langle\rho(t,x)\rangle_\psi$
is nonzero for some $t$. This ergodic result shows that the spatial and
temporal averages of energy densities obey related constraints. 

As a second illustration, consider (IV) in the case where $H$ has a
discrete spectrum $\{E_n\}$ with corresponding orthonormal eigenfunctions
$\{\phi_n\}$, and satisfying the integrability condition on $\mu_x$ (for
example, $H$ might be semibounded). Then, in the case $c=0$, 
\begin{eqnarray}
\langle h_x(g^2)\rangle_\psi &\ge& -\int\frac{du}{2\pi}\, |\widehat{g}(u)|^2
\sum_{E_n\le u} |\phi_n(x)|^2(\hbar u-E_n) \nonumber\\
&=& -\sum_{n} \alpha_n |\phi_n(x)|^2
\end{eqnarray}
where 
\begin{equation}
\alpha_n = \int_{E_n}^\infty du\, |\widehat{g}(u)|^2(\hbar u-E_n)\,.
\end{equation}
These formulae may be used to compare the relative ease of obtaining
negative energy densities at different spatial locations. For example,
the eigenfunctions $\phi_n$ of the harmonic oscillator
$H=p^2/(2m)+\frac{1}{2}m\omega^2 x^2$ on $L^2(\RR)$ obey the following
bounds (cf.\ the Appendix to Sec.\ V.3 in \cite{ReedSimon1}):  For any $j\in\NN_0$
there exists $c_j>0$ and $r_j\in\NN_0$ such that 
\begin{equation}
\sup_{x\in\RR} |(1+x^j) \phi_n(x)| \le c_j (1+n)^{r_j}
\end{equation}
for all $n\in\NN_0$. Thus, for all normalised $\psi\in\Sch(\RR)$ (in fact, for all
$\psi$ in a considerably larger domain) we have
\begin{equation}
\langle h_x(g^2)\rangle_\psi \ge -\frac{c_j}{1+|x|^j}\sum_{n=0}^\infty
\alpha_n (1+n)^{r_j}\,.
\end{equation}
In this case, it is clear that---for a fixed sampling function $g$---the $\alpha_n$
form a rapidly decaying sequence and so the sum converges for any $j$.
Thus we have shown that the state-independent bound on energy density is
itself a rapidly decaying function of $x$. It is therefore generally
easier to maintain negative energy densities near the classical
equilibrium point rather than far away. 

Finally, consider (III) for the case $H=-id/dx$ on $\HH=L^2(\RR)$ and a
particle of mass $m$. In this instance, the spaces $\HH_k$ coincide with
the Sobolev spaces $W^{k,2}(\RR)$ and Sobolev's lemma permits us to
take $k>1/2$. Then the dynamical evolution amounts to spatial translation and the averaged
Hamiltonian density is related to the spatially averaged probability
flux by
\begin{equation}
\langle h_x(f)\rangle_\psi = m\,j_\psi(\tilde{f}_x)
\end{equation}
where $\tilde{f}_x(t)=f(x-t)$. Moreover, the measure $\mu_x$ is easily
seen to be given by $\mu_x(\Delta)=|\Delta|/(2\pi\hbar)$, where
$|\cdot|$ denotes the usual Lebesgue measure. Then the second inequality
in (III) may easily be checked to reproduce the flux
inequality~\eqref{eq:fluxQI} for all $\psi\in P_{[0,\infty)}W^{k+1,2}(\RR)$.
\\[6pt]
The remainder of this section is devoted to the proof of statements (III) and (IV).
These assertions are based upon two facts, which will be proved below. 
First, for any $c\in\RR$ and normalised $\psi\in \HH_{k+1}$, one may show that
\begin{equation}
\langle h_x(f)\rangle_\psi - c\langle \rho_x(f)\rangle_\psi 
= \int\frac{d\epsilon}{2\pi\hbar} (\epsilon-c)\left|
\ip{\psi}{p_k(H)\widehat{g}(\hbar^{-1}[\epsilon\II-H])\eta_x} \right|^2\,.
\label{eq:fact1}
\end{equation}
Second, if $\Delta$ is a Borel set then
\begin{equation}
\left|
\ip{\psi}{p_k(H)\widehat{g}(\hbar^{-1}[\epsilon\II-H])\eta_x} \right|^2
\le \int_\Delta d\mu_x(E)\,
\left|\widehat{g}([\epsilon-E]/\hbar)\right|^2\,.
\label{eq:fact2}
\end{equation}
for all normalised $\psi\in P_\Delta\HH$. Putting these together, we obtain
\begin{eqnarray}
\langle h_x(f)\rangle_\psi - c\langle \rho_x(f)\rangle_\psi &\ge&
\int \frac{d\epsilon}{2\pi\hbar} \{\epsilon-c\}_- \int_\Delta d\mu_x(E)\,
\left|\widehat{g}([\epsilon-E]/\hbar)\right|^2 \nonumber\\
&=& \int\frac{du}{2\pi}\, |\widehat{g}(u)|^2 \int_\Delta d\mu_x(E)\,
\{\hbar u+E-c\}_-  \nonumber\\
&=&- \int\frac{du}{2\pi}\, |\widehat{g}(u)|^2 \int_\Delta d\mu_x(E)\,
\{\hbar u+c-E\}_+\,, 
\label{eq:fact_calc}
\end{eqnarray}
for all normalised $\psi\in P_\Delta\HH$, where 
we have made the change of variables $u\to -u$
and exploited the fact that $|\widehat{g}(u)|^2$ is even (because $g$ is
real-valued). The interchange of variables employed in the first step is
justified provided the inner integral in the last line
of~\eqref{eq:fact_calc} is polynomially bounded in $u$. 

{}To obtain the second inequality in (III), we set $c=a$ and
$\Delta=[a,b]$ and observe that the inner integral in
Eq.~\eqref{eq:fact_calc} reduces to $Q_-(u)$
and is polynomially bounded because $\mu_x$ is. The inequality clearly remains
true for $\psi\in P_{[a,\infty)}\HH_{k+1}$ with $\Delta=[a,\infty)$. 
To obtain (IV),
we set $\Delta=\RR$ and observe that the integrability condition 
$\int_{\RR^-} d\mu_x(E) (1+|E|)< \infty$ and polynomial boundedness of
$\mu_x$ guarantee that $S(H-c\II,u)$ exists and is polynomially bounded. 
Inequality~\eqref{eq:IVstmnt} follows from the
above on observing that
$\int d\mu^{(H)}_x(E)\, F(E-c)=\int d\mu^{(H-c\II)}_x(E)\, F(E)$. 

To obtain the first inequality in (III) and the dual statement to (IV),
one argues in an analogous fashion from the calculation
\begin{eqnarray}
\langle h_x(f)\rangle_\psi - c\langle \rho_x(f)\rangle_\psi &\le&
\int \frac{d\epsilon}{2\pi\hbar} \{\epsilon-c\}_+ \int_\Delta d\mu_x(E)\,
\left|\widehat{g}([\epsilon-E]/\hbar)\right|^2 \nonumber\\
&=& \int\frac{du}{2\pi}\, |\widehat{g}(u)|^2 \int_\Delta d\mu_x(E)\,
\{\hbar u+E-c\}_+\,, 
\end{eqnarray}
which holds for all normalised $\psi\in P_\Delta\HH$. 

\vspace{0.1cm}
It remains to prove the two facts presented as Eqs.~\eqref{eq:fact1}
and~\eqref{eq:fact2} above. First, observe that for any normalised $\psi\in\HH_k$, $\langle
\rho_x(f)\rangle_\psi$ may be expressed as 
\begin{equation}
\langle \rho_x(f)\rangle_\psi = \int dt\, f(t) 
\int d\ip{\psi}{P_E\eta_x}\int d\ip{\eta_x}{P_{E'}\psi}
e^{i(E-E')t/\hbar}p_k(E)p_k(E') 
\end{equation}
by the functional calculus. Performing the $t$ integral first 
(which is legitimate since $f$ is
smooth and compactly supported) we obtain 
\begin{equation}
\langle \rho_x(f)\rangle_\psi = 
\int d\ip{\psi}{P_E\eta_x}\int d\ip{\eta_x}{P_{E'}\psi}
\widehat{f}([E'-E]/\hbar)p_k(E)p_k(E')\,.
\label{eq:rhof}
\end{equation}
Since $f=g^2$, the convolution theorem may be used to write
\begin{equation}
\widehat{f}([E'-E]/\hbar) = \int \frac{d\epsilon}{2\pi\hbar} \,\widehat{g}([E'-\epsilon]/\hbar)
\overline{\widehat{g}([E-\epsilon]/\hbar)} 
\end{equation}
using the fact that
$\overline{\widehat{g}(\lambda)}=\widehat{g}(-\lambda)$ since $g$ is
real-valued. Substituting in~(\ref{eq:rhof}), 
and again rearranging the order of integration, we obtain 
\begin{eqnarray}
\langle \rho_x(f)\rangle_\psi &=& \int\frac{d\epsilon}{2\pi\hbar} 
\int d\ip{\psi}{P_E\eta_x}\int d\ip{\eta_x}{P_{E'}\psi}
\widehat{g}([E'-\epsilon]/\hbar)
\overline{\widehat{g}([E-\epsilon]/\hbar)}p_k(E)p_k(E')\nonumber\\
&=& \int\frac{d\epsilon}{2\pi\hbar} \left| \int d\ip{\psi}{P_E\eta_x}
p_k(E)\overline{\widehat{g}([E-\epsilon]/\hbar)}\right|^2\nonumber\\
&=& \int\frac{d\epsilon}{2\pi\hbar} \left|
\ip{\psi}{p_k(H)\widehat{g}(\hbar^{-1}[\epsilon\II-H])\eta_x} \right|^2
\label{eq:fact1.5}
\end{eqnarray}
To treat $\langle h_x(f)\rangle_\psi$ for normalised $\psi\in\HH_{k+1}$, we write
\begin{equation}
\langle h_x(f)\rangle_\psi = \frac{1}{2}\int dt\, f(t)
\int d\ip{\psi}{P_E\eta_x}\int d\ip{\eta_x}{P_{E'}\psi}
e^{i(E-E')t/\hbar}(E+E')p_k(E)p_k(E')\,,
\end{equation}
by functional calculus and use the identity
\begin{equation}
\frac{(E+E')}{2}\widehat{f}([E'-E]/\hbar) = \int \frac{d\epsilon}{2\pi\hbar} \epsilon
\widehat{g}([E'-\epsilon]/\hbar)
\overline{\widehat{g}([E-\epsilon]/\hbar}) 
\end{equation}
in place of the convolution theorem. (See~\cite{FVdirac} and~\cite{FewsterMistry}
for proofs of this identity.)

By a derivation analogous to that used for $\langle\rho_x(f)\rangle_\psi$ we
then obtain
\begin{equation}
\langle h_x(f)\rangle_\psi =
\int\frac{d\epsilon}{2\pi\hbar} \epsilon\left|
\ip{\psi}{p_k(H)\widehat{g}(\hbar^{-1}[\epsilon\II-H])\eta_x} \right|^2
\end{equation}
and Eq.~\eqref{eq:fact1} follows from this equation
and~\eqref{eq:fact1.5}.

\vspace{0.1cm}
The second assertion, Eq.~\eqref{eq:fact2}, is proved by noting that
\begin{equation}
\left| \ip{\psi}{p_k(H)\widehat{g}(\hbar^{-1}[\epsilon\II-H])\eta_x} \right|^2
\le \|P_\Delta p_k(H)\widehat{g}(\hbar^{-1}[\epsilon\II-H])\eta_x\|^2
\end{equation}
using $\psi=P_\Delta\psi$ and the Cauchy--Schwarz inequality (with
$\|\psi\|=1$). The right-hand side may be written
\begin{equation}
\int_\Delta d\ip{\eta_x}{P_E\eta_x}
p_k(E)^2|\widehat{g}([\epsilon-E]/\hbar)|^2 =
\int_\Delta d\mu_x(E) |\widehat{g}([\epsilon-E]/\hbar)|^2 
\end{equation}
which completes the derivation of Eq.~\eqref{eq:fact2}.

\section{Numerical Details} \label{sect:numerics}

In this section we provide more details on the numerical methods
employed in Sect.~\ref{sect:backflow} and discuss rigorous error
estimates on the numerical errors. The basic numerical method is easily explained
(see, e.g., Sec 4.1 of~\cite{Atkinson} or Chapter 4
of~\cite{DelvesMohamed}). Suppose $T$ is an integral operator on
$L^2(\RR^+,dk)$ with kernel $G$, i.e.,
\begin{equation}
(T\psi)(k)=\int_0^\infty dk' G(k,k')\hat{\psi}(k')\,.
\end{equation}
To handle this numerically, we first truncate the kernel to $[0,K]\times
[0,K]$ for some $K>0$, which amounts to studying a compression $T_K$
of $T$. Provided that the required properties of $T$ are, for
sufficiently large $K$, well-approximated by the corresponding
properties of $T_K$ restricted to $L^2(0,K)$, we proceed to approximate
this restricted operator by a matrix. To do this, we suppose that 
$(\xi_j)_{j=0}^N$
and $(w_j)_{j=0}^N$ are the nodes and weights for a suitable quadrature method on
$[0,K]$, and define the $(N+1)$-square matrix $A=(A_{jk})_{j,k=0}^N$ with $(j,k)$ entry 
$A_{jk}=w_j^{1/2}w_k^{1/2}G_{\lambda,K}(\xi_j,\xi_k)$. The relevant
computations are performed on $A$ and, if $N$ and $K$ are sufficiently
large, this will provide a numerical approximation to the required
quantity. 

This technique was applied to the Bracken--Melloy kernel as described in
Sect.~\ref{sect:fluxQI}. In that case, we were unable to derive useful
error estimates. However, the operator norm calculations of
Sect.~\ref{sect:fluxnumerics} are more controlled, as we now describe.
The problem is to estimate the squared operator norms of the family of integral operators
$T_\lambda$ ($\lambda>0)$ defined in the above fashion\footnote{In
Sect.~\ref{sect:fluxnumerics} the operators were defined on
$L^2(\RR^+,dk/(2\pi))$. Here, we absorb the factor of $(2\pi)^{-1}$
into the kernel, which leaves the spectral data and operator norms unchanged.}
 with kernel
\begin{equation}
G_{\lambda}(k,k')=\frac{1}{2\pi}\sqrt{k}\hat{g}_{\lambda}(-k-k')\,.
\end{equation}
Now the compressions $T_{K,\lambda}$ converge to $T_\lambda$ in the
Hilbert--Schmidt norm, and therefore in operator norm, as $K\to\infty$. 
That the matrix approximations have operator norms converging to
$\|T_{K,\lambda}\|$ as $N\to\infty$ is a consequence of the convergence
of the quadrature formula to the integral for continuous functions. Thus
our technique may be validly applied to this problem and it remains to
control the errors inherent in the scheme for finite $N$ and $K$. In
general we have analytical control of the truncation errors
(parametrised by $K$) but not the discretisation errors; we are,
however, able to observe apparent convergence to machine precision in
most cases. As a first step in our analysis of the truncation errors, we 
eliminate the parameter $\lambda$ from our
considerations: $G_\lambda$ and the dilation of $G_1$ by a factor of
$\lambda$ differ only by a constant factor of $\lambda$, so if
truncation of $G_1$ at $K$ has a relative error of $\eps$ (in either
Hilbert-Schmidt or operator norm) then truncation of $G_\lambda$ at $\lambda K$
also has relative error $\eps$, and the associated quadrature matrices
are identical apart from a factor of $\lambda$. From now on, we will use the 
value $\lambda=1$ only, and write $T_K$ for $T_{K,1}$, etc.

In order to estimate truncation errors, the following observations are
useful. We wish to integrate $|G|^2$ over the region 
$[0,\infty)\times[0,\infty)\setminus[0,K]\times[0,K]$. By symmetry we can
integrate $(|G(k,k')|^2+|G(k',k)|^2)/2$ over the same region to obtain the
same result. This has the advantage that
\begin{equation}
\frac{|G(k,k')|^2+|G(k',k)|^2}{8\pi^2}=
  \frac{(k+k')|\hat{g}(-k-k')|^2}{8\pi^2}
\end{equation}
which depends only on $k+k'$. We can exploit this by using the following
decomposition of the quadrant $[0,\infty)\times[0,\infty)$
\begin{center}
  \begin{pspicture}(-0.5,-0.5)(5,5)
    \psaxes[labels=none,ticks=none]{<->}(0,0)(0,5)(5,0)
    \psline[linestyle=dotted](2,0)(2,2)(0,2)
    \psline[linestyle=dotted](4,0)(0,4)
    \rput(2,-0.3){$K$}
    \rput(4,-0.3){$2K$}
    \rput(-0.3,2){$K$}
    \rput(-0.3,4){$2K$}
    \rput(2.7,0.7){$R_1$}
    \rput(0.7,2.7){$R_2$}
    \rput(3,3){$R_3$}
  \end{pspicture}
\end{center}
We wish to evaluate the integral over $R_1\cup R_2\cup R_3$, and by
symmetry the integrals over $R_1$ and $R_2$ are equal. Changing into a
$(k+k',k-k')$ coordinate system we see that the required integral is
\begin{equation}\label{eqn:truncation-error}
  \frac{1}{4\pi^2}\int_{K}^{2K}u(u-K)|\hat{g}(-u)|^2du+
  \frac{1}{8\pi^2}\int_{2K}^{\infty}u^2|\hat{g}(-u)|^2du\,.
\end{equation}

We now consider the four sampling functions used in
Sect.~\ref{sect:fluxnumerics}. Starting with the Gaussian kernel
\begin{equation}
\hat{g}(k)=\sqrt{2}\pi^{1/4}e^{-k^2/2}\,,
\end{equation}
the Hilbert--Schmidt norm of $T$ can be found by substituting $K=0$ in
\eqref{eqn:truncation-error} and evaluating the integral to give 
\begin{equation}
\|T\|_{\hs}=\frac{1}{4\sqrt{\pi}}
\end{equation}
which is of course an upper bound for the operator norm.

For more precise results, we turn to the quadrature method described above.
For this kernel, the integrals in \eqref{eqn:truncation-error} can be 
evaluated explicitly to give a relative error in the Hilbert-Schmidt norm of
\begin{equation}
\frac{\|T-T_K\|_\hs}{\|T\|_\hs}=(1+\erf(2K)-2\erf(K))^{1/2}\,.
\end{equation}
It can be numerically verified that the relative error falls below 
$\eps=0.5\times10^{-10}$ (for ten-digit precision) at approximately 
$K=6.756$ (this calculation requires about 25-digit precision).

The computations were performed in Maple~8 using $c$-panel repeated Clenshaw-Curtis quadrature
(see Section 2.4.4 of \cite{DelvesMohamed}) on the interval $[0,6.9]$ 
and Maple's NAG-based \texttt{SingularValues} routine. Using $33$, $65$
or $129$ samples with $c=1,2$ gives in each case the same results for
the first largest two singular values: 
$\sigma_1=0.1399331442$, $\sigma_2=0.0175697912$ to 10 figures. 
Notice that the second singular value is very much smaller than the
first, which means that the matrices, and hence $T$, can be well
approximated by operators of rank 1. This is consistent with the
operator norm, computed here to be $.1399331442$, being close to the
Hilbert-Schmidt norm, $1/(4\sqrt{\pi})=.1410473959\dots$. This
similarity finally justifies our use of truncation constants based on the 
relative error in the Hilbert-Schmidt norm: the calculated value of the
operator norm is certainly no larger than the true value, since it is
the norm of a compression of $T$, so we have
\begin{eqnarray}
\frac{\|T-T_K\|}{\|T\|}&\leq&\frac{\|T-T_K\|_\hs}{\|T\|_\hs}\frac{\|T\|_\hs}{\|T\|}\nonumber\\
 &\leq& .4873572016\times10^{-10}\frac{.1410473959}{.1399331442}
  =.4912379017\times10^{-10}
\end{eqnarray}
which is still less than the target figure of $0.5\times10^{-10}$.

The next kernel of interest is the squared Lorentzian; however, in this
case $T$ is a rank-1 operator so the Hilbert--Schmidt and operator norms
coincide and there is no need for numerical investigation. This leaves
the two compactly supported kernels. In the truncated cosine case, the
same techniques as above lead to an error estimate of the order of $2\%$
relative error in the Hilbert--Schmidt norm for $K=1100$. As this is a
rather weak estimate, we suppress the details; the numerical estimate of
the squared operator norm (for $K=1100$, $N=1024$) is given in Sect.~\ref{sect:fluxnumerics}. 

Our last example is the smoothed truncated cosine,
defined by
\begin{equation}
\hat{g}(k)=\frac{2\sqrt{3}\pi^2}{3}\frac{\sin(k)}{(\pi^2-k^2)k}\,.
\end{equation}
The relatively slow decay of this function makes the precision obtained
in the Gaussian example impractical, but we can obtain results to at
least four significant figures. In fact the numerical results appear to be much more
precise than would be suggested by this error estimate. 

Maple is able explicitly to evaluate the integrals in 
\eqref{eqn:truncation-error} to give a rather complicated formula involving the 
Si and Ci special functions, and from this to give the asymptotic formula
\begin{equation}
\frac{5\pi}{16K^3}+o\left(\frac{1}{K^4}\right)
\end{equation}
for the relative error in the Hilbert-Schmidt norm. Using only the
leading term, we can predict that truncation at about $K=732.3$ should give a
Hilbert-Schmidt norm relative error less than $0.5\times10^{-4}$;
numerical investigation of the exact formula near this point confirms
this value. Proceeding in the same way as for the Gaussian kernel
but this time using the faster numerical engine in Matlab~6 to calculate
the singular values, we obtain the following results ($N+1$ samples, $c$ panels):
\[
\begin{array}{r|l|l|l}
N\ \ & c& \sigma_1    & \sigma_2      \\\hline
 256 & 1 & .3536210415 & .0733902393 \\
 256 & 2 & .3536211355 & .0733900951 \\
 512 & 1 & .3536210388 & .0733902259 \\
 512 & 2 & .3536210388 & .0733902259 \\
1024 & 1 & .3536210388 & .0733902259 \\
1024 & 2 & .3536210388 & .0733902259 \\
\end{array}
\]

Once again, the fact that the second singular value is considerably
smaller than the first can be used after the fact to justify the use of
relative errors in the Hilbert-Schmidt norm (rather than in the operator norm) in
choosing the truncation constant.

Although the error analysis only allows us to be confident of the first
four figures, it seems likely that this figure for the operator norm is
considerably more accurate than that. Doubling the truncation constant
and using $2049$ points, again with $1$ and $2$ panels, gives exactly the 
same results to ten figures as the two $1025$-point methods above.

The last set of calculations reported in Sect.~\ref{sect:fluxnumerics}
concern the unbounded operator $J$ of Eq.~\eqref{eq:Jdef}. Here we have
not succeeded in obtaining usable estimates of the errors introduced by
truncation to $[0,K]$. However, it is nonetheless true that
$\inf \sigma(J_K)\to\inf \sigma(J)$ as a consequence of the following
arguments. 

\vspace{0.2cm}
{\noindent\bf Proposition} {\em Suppose $k$ is an absolutely bounded kernel on $L^2(0,\infty)$, and
let $w$ be a measurable function on $(0,\infty)$ (with respect to
Lebesgue measure). Let
\begin{equation}
D=\{f\in L^2(0,\infty) :wf\in L^2(0,\infty)\}\,.
\end{equation}
Suppose $(w(x)+w(y))k(x,y)$ is a Hermitian function of $x$ and $y$, and 
define an operator with domain $D$ by
\begin{equation}
(Tf)(x)=\int_0^\infty (w(x)+w(y))k(x,y)f(y)dy
\end{equation}
and assume that $T$ is bounded below. For $K>0$ define the truncated operator
\begin{equation}
(T_Kf)(x)=\int_0^K(w(x)+w(y))k(x,y)f(y)dy\,.
\end{equation}
Then
\begin{equation}
\lim_{K\to\infty}\inf\sigma(T_K)=\inf\sigma(T)\,.
\end{equation}}

\noindent\textbf{Proof:} $T_K$ is a compression of $T$ so $\inf\sigma(T_K)\geq\inf\sigma(T)$ for
all $K$. For any $\eps>0$ we can choose $f\in D$ with $\|f\|=1$ and 
such that
$\ip{f}{Tf}<\inf\sigma(T)+\eps/2$. Now
\begin{equation}
\ip{f}{Tf}=\int_0^\infty
  \left(\int_0^\infty(w(x)+w(y))k(x,y)f(y)dy\right)\overline{f(x)}dx
\end{equation}
and the integrand here is in $L^2((0,\infty)\times(0,\infty))$ by the lemma.
It now follows from Lebesgue's dominated convergence theorem and
Fubini's Theorem that, provided the above repeated integral can be
interpreted as an integral on the measure space $(0,\infty)\times (0,\infty)$,
\begin{equation}
\ip{f}{Tf}=\lim_{K\to\infty}\int_0^K
  \left(\int_0^K(w(x)+w(y))k(x,y)f(y)dy\right)\overline{f(x)}dx\,.
\end{equation}
If we let $f_K(x)=f(x)\chi_{(0,K)}(x)$ then this tells us that
$\ip{f_K}{T_Kf_K}\to\ip{f}{Tf}$; we lso have $\|f_K\|\to\|f\|=1$ as
$K\to\infty$, so $\ip{f_K}{T_Kf_K}/\|f_K\|^2\to\ip{f}{Tf}$ as
$K\to\infty$. In particular, for sufficiently large $K$ we have 
\begin{equation}
\frac{\ip{f_K}{T_Kf_K}}{\|f_K\|^2}<\ip{f}{Tf}+\frac{\eps}{2}<\inf\sigma(T)+\eps
\end{equation}
which implies that $\inf\sigma(T_K)<\inf\sigma(T)+\eps$.
In combination with the earlier inequality $\inf\sigma(T_K)\geq\inf\sigma(T)$,
this establishes the result.

It remains to justify the treatment of the repeated integral as an
integral on a product measure space. 

\vspace{0.2cm}
{\noindent\bf Lemma} {\em
In the notation of the above theorem, for any $f,g\in D$, the repeated integral in the bilinear form
\begin{equation}
\int_0^\infty\left(\int_0^\infty(w(x)+w(y))k(x,y)f(x)g(y)dy\right)dx
\end{equation}
is absolutely convergent, and so can be interpreted as the integral of a
function in $L^2((0,\infty)\times (0,\infty))$.}

\noindent \textbf{Proof:} We calculate
\begin{align}
    &\phantom{=}\int_0^\infty\left(\int_0^\infty|(w(x)+w(y))k(x,y)f(x)g(y)|dy\right)dx \nonumber\\
    &\leq \int_0^\infty\left(\int_0^\infty(|w(x)|+|w(y)|)|k(x,y)|\,|f(x)|\,|g(y)|dy\right)dx \nonumber\\
    &=    \int_0^\infty\left(|f(x)w(x)|\int_0^\infty|k(x,y)|\,|g(y)|dy\right)dx+ \nonumber\\
    &\phantom{=} \int_0^\infty\left(|f(x)|\int_0^\infty|k(x,y)|\,|w(y)g(y)|dy\right)dx\,.
\end{align}
In the first term, the inner integral defines an $L^2$ function of $x$
since $k$ is an absolutely bounded kernel, and $f(x)w(x)$ is an $L^2$
function of $x$ since $f\in D$. The first term is therefore finite by the 
Cauchy-Schwarz inequality. The second term is finite by similar reasoning: 
$w(y)g(y)$ is an $L^2$ function of $y$ since $g\in D$, so the inner
integral define an $L^2$ function of $x$, and $f$ is $L^2$ by hypothesis.
The final conclusion now follows from Tonelli's Theorem.

\section{Conclusion}\label{sect:conclusion}

The main focus of this paper has been to draw attention to links between
the failure of pointwise energy conditions in quantum field theory and a
range of similar situations in quantum mechanics. In addition we have
seen that there are links at the technical level between the QIs
developed in quantum field theory and those obtained here in the quantum
mechanical setting. In addition, we have made contact with the ideas and
methods of Weyl--Wigner quantisation and sharp G{\aa}rding inequalities.
In conclusion, we briefly summarise the new results we have
obtained along the way. 

First, we have seen that the backflow phenomenon is limited in space (as
well as in time~\cite{BrackenMelloy}) as shown by our flux
QI~\eqref{eq:fluxQI}. In particular, the magnitude of the negative flux
times the square of its spatial extent is bounded above for all
right-moving states in $\Rr$. We have also provided an improved
numerical estimate of Bracken and Melloy's backflow constant, and also
given numerical evidence to support the conjecture that our flux QI is
generally within an order of magnitude of the optimal bound (i.e., the
infimum of the spectrum of $J$, given by Eq.~\eqref{eq:Jdef}). 

Second, we have shown that similar phenomena occur for densities of
observables obtained via Weyl quantisation. This is a consequence of the
indefinite sign of the Wigner function, and therefore an expression of
the uncertainty principle. Moreover, for observables which are second order
(or less) in
momentum, we have seen that sharp G{\aa}rding inequalities entail the
existence of kinematic quantum inequalities. We have also obtained
explicit bounds in the case of Schr\"odinger operators with smooth
potentials. 

Finally, for general quantum mechanical systems describing dynamics on a
topological measure space, we have shown that the time-averaged energy density obeys
dynamical quantum inequalities (evolution being generated by the spatial
integral of the energy density). For the 1-dimensional harmonic oscillator, we saw that the QI
bound (for a given sampling function) is a Schwartz-class function: it
becomes rapidly much harder to create sustained negative energy
densities away from the classical equilibrium point. 
Moreover, we have seen that a bound on the spectral behaviour of the Hamiltonian on 
the negative spectral axis, expressed by the integrability condition on $\mu_x$
in (IV), already leads to dynamical QIs. This integrability condition can be
viewed as a condition on the global dynamical stability of a quantum system,
much in the sense of quantum systems in thermal equilibrium, where the
spectral weight of the generator of the time-evolution (the Liouvillian) is
exponentially suppressed on the negative half-axis (cf.\ Prop.\ 5.3.14 in
\cite{BratRob2}). This indicates again the link between (thermo)dynamical
stability and dynamical QIs which was established in \cite{passivity} and which
originally motivated the introduction of QIs in \cite{Ford78}.

All these findings corroborate the intimate connection between QIs and the
fundamental principles of quantum mechanics: the uncertainty 
principle and dynamical stability. \\
${}$\\
{\noindent\em Acknowledgments:} CJF thanks I\~{n}igo Egusquiza for
bringing ref.~\cite{BrackenMelloy} to his attention. The work of CJF was
assisted by EPSRC Grant GR/R25019/01 to the University of York; RV also
thanks the EPSRC for support received under this grant during a visit to
York. Numerical calculations were partly conducted using the White Rose
Grid node hosted at the University of York. Some
of this work was conducted at the Erwin Schr\"odinger Institute, Vienna,
during the programme on Quantum Field Theory in Curved Spacetime which
took place in July--August 2002; CJF and RV
thank the organisers of this programme and the institute for its hospitality.
Particular inspiration was drawn from the Hotel Sacher.

\end{document}